\documentclass[aps,twocolumn,amssymb,superscriptaddress,nofootinbib,longbibliography,pra]{revtex4-2}

\usepackage{graphicx}
\usepackage{color}
\usepackage{braket}
\usepackage{verbatim}
\usepackage{amsmath}
\usepackage{mathtools}
\usepackage{braket}
\usepackage{float}
\usepackage{hyperref}
\usepackage{dsfont}
\usepackage{lipsum}
\usepackage{commath}
\usepackage{amsfonts}
\usepackage{tabularx,ragged2e}

\usepackage[ruled,vlined,linesnumbered]{algorithm2e}
\SetCommentSty{mycommfont}

\makeatletter
\renewcommand*\env@matrix[1][*\c@MaxMatrixCols c]{
  \hskip -\arraycolsep
  \let\@ifnextchar\new@ifnextchar
  \array{#1}}
\makeatother

\begin{document}
\title{Purification of photonic graph states}

\author{Matthias C. L{\"o}bl}
\email{mlo@sparrowquantum.com}
\affiliation{Sparrow Quantum, Nordre Fasanvej 215, DK-2000 Frederiksberg, Denmark}
\author{Aliki A. Capatos}
\affiliation{NNF Quantum Computing Programme, Niels Bohr Institute, University of Copenhagen, Blegdamsvej 17, DK-2100 Copenhagen {\O}, Denmark}
\affiliation{Quantum Engineering Centre for Doctoral Training, University of Bristol, Bristol, United Kingdom}
\author{Ming Lai Chan}
\affiliation{Sparrow Quantum, Nordre Fasanvej 215, DK-2000 Frederiksberg, Denmark}
\affiliation{Center for Hybrid Quantum Networks (Hy-Q), The Niels Bohr Institute, University~of~Copenhagen,  DK-2100  Copenhagen~{\O}, Denmark}
\author{Peter Lodahl}
\affiliation{Sparrow Quantum, Nordre Fasanvej 215, DK-2000 Frederiksberg, Denmark}
\affiliation{Center for Hybrid Quantum Networks (Hy-Q), The Niels Bohr Institute, University~of~Copenhagen,  DK-2100  Copenhagen~{\O}, Denmark}
\author{Anders S{\o}ndberg S{\o}rensen}
\affiliation{Center for Hybrid Quantum Networks (Hy-Q), The Niels Bohr Institute, University~of~Copenhagen,  DK-2100  Copenhagen~{\O}, Denmark}
\author{Stefano Paesani}
\affiliation{NNF Quantum Computing Programme, Niels Bohr Institute, University of Copenhagen, Blegdamsvej 17, DK-2100 Copenhagen {\O}, Denmark}

\begin{abstract}
Graph states constitute the main building block for quantum computing with photons. Quantum emitters with a hosted spin can deterministically generate photonic graph states, strongly lowering the overhead of multiplexing highly probabilistic linear-optics graph state generation. However, they generally suffer from various noise sources, resulting in reduced fidelities of the produced states. To mitigate this issue, we develop purification schemes for entangled photonic states. We first develop purification schemes to purify arbitrary photonic GHZ and other CSS states, which we generalize to all photonic graph states and stabilizer states. The proposed purification schemes have a high success probability of up to $1/2$ and require only linear optics and photon detectors. We optimize cascaded purification schemes for various graph states, taking into account phenomenological Pauli errors or physical noise in time-bin-encoded graph state generation with quantum emitters.
\end{abstract}

\maketitle

\section{Introduction}
Entangled photonic states are the key resource for measurement- and fusion-based quantum computing~\cite{Raussendorf2001, Bartolucci2021}, as well as one-way quantum repeaters~\cite{Azuma2015}. All-linear-optic schemes can generate entanglement only probabilistically and the success probability decays exponentially with the number of qubits~\cite{Gubarev2020, Cao2025, Zatsch2025}. In contrast, solid-state quantum emitters with a spin enable deterministic generation of large states with a linear entanglement structure~\cite{Lindner2009}. However, only limited fidelities have been reached so far, even for small graph states~\cite{Schwartz2016, Meng2023b, Cogan2023, Thomas2022, Huet2024, Huet2026}. An issue for solid-state systems is that intrinsic noise sources, such as phonons or nuclear spin noise~\cite{Tiurev2021}, may set a limit for the achievable fidelities. Furthermore, fidelities decrease further when fusions~\cite{Lee2023, Lobl2024} or multiple interacting spins~\cite{Li2022} are used to build more complex resource states, since every additional operation is subject to noise, resulting in an accumulation of errors. On the application side, low error rates are required for fusion-based quantum computing~\cite{Bartolucci2021, Chan2025}. A promising route to achieve sufficiently low error rates, i.e. high enough fidelities, is the use of purification schemes.

Here, we propose and optimize several purification schemes for photonic graph states, going beyond previous schemes for purifying two-qubit Bell pairs~\cite{Pan2001, Pan2003, Hu2021, Yu2025} or single photons~\cite{Faurby2024,Somhorst2026}. In contrast to previous purification schemes for multipartite states~\cite{Dur2003, Aschauer2005, Goyal2006, Kruszynska2006, Wang2026}, our schemes do not use deterministic photon-photon entangling gates, since such gates are challenging to implement. Instead, our purification schemes require only linear optics and photon-number-resolving detectors, where type-I fusions~\cite{Browne2005} are transversally applied. We first provide a purification scheme that can be applied to all photonic Calderbank-Shor-Steane (CSS) states~\cite{Calderbank1996, Steane1996}, including Greenberger-Horne-Zeilinger (GHZ) states. Furthermore, this purification scheme can be applied to all two-colorable (bicolorable) graph states, as these are locally equivalent to CSS states~\cite{Chen2004}. The purification scheme takes two imperfect copies of the desired CSS state, sends them to a linear optics setup, and destructively measures half of the qubits. The output state is kept if a measurement-dependent heralding criterion is met, resulting in a state with higher fidelity. Finally, we provide a more general purification scheme for all graph states. This scheme also works via linear optics and uses ancillary photonic graph states. It can be applied to all stabilizer states as they are locally equivalent to graph states~\cite{Nest2004}.

For several entangled states, we numerically find the optimal order of cascaded purification schemes to achieve the highest fidelity. The best order turns out to depend on the exact noise parameters for which we consider phenomenological bit- and phase-flip errors, depolarization noise, as well as physical noise sources in an emitter-based protocol for generating time-bin-encoded graph states~\cite{Tiurev2021, Tiurev2021b}. We find that purification can strongly reduce error rates. For example, for a three qubit (four-qubit) GHZ state with fidelity $F\approx0.86$ ($F\approx0.81$) due to depolarization noise, we find that it can be purified to a fidelity of $F=0.988$ ($F=0.981$) in two cascaded purification rounds with an overall success probability of $1/8$ ($1/16$). Finally, we analyze the success probability of the purification schemes and outline strategies to minimize the multiplexing overhead.

The paper is structured as follows. In Section~\ref{sec_setup}, we first introduce type-I fusion and a high-level setup for entangled state generation with quantum emitters and built-in purification. In Section~\ref{sec_ghz}, we analyze the purification of GHZ states, considering the states in the Schrödinger picture. Readers familiar with the stabilizer formalism may jump directly to Sections~\ref{sec_graph_css_def},~\ref{sec_css},~\ref{sec_all_graph} where we consider the purification of more general stabilizer states, in particular graph and CSS states. In Section~\ref{sec_time_bin} we describe the generation and purification of time-bin-encoded states, and in Section~\ref{sec_numerics} we numerically analyze the performance of purification protocols under phenomenological and physical noise inherent to quantum emitters. Finally, we discuss the multiplexing overhead for building deterministic sources of purified resource states in Section~\ref{sec_multiplex}.

\section{Physical setup}
\label{sec_setup}
\subsection{Type-I fusion}
\label{sec_type1}
The key component of the purification protocols investigated here is transversal type-I fusion. In a type-I fusion, two photons $A, B$ are sent to two different modes of a linear optics setup, and the detection of exactly one photon in one of the output arms implements the desired measurement (fusion success)~\cite{Browne2005, Gimeno2016}. Detecting two or zero photons corresponds to fusion failure. Throughout the article, we mostly refer to type-I fusion for polarization-encoded qubits (linear $H,V$ polarization), but equivalent setups exist for various encodings such as path (see Appendix~\ref{sec_dual_rail}) or time-bin encoding (see Section~\ref{sec_time_bin}). For polarization-encoded qubits, type-I fusion consists of a polarizing beam splitter (PBS), followed by a photon-number-resolving detector that measures the photon in the diagonal basis corresponding to $H+V, H-V$ polarization. This setup is shown in Fig.~\ref{fig_setups}(a), where it forms a part of a setup for purifying a three-qubit GHZ state. As seen in the figure, type-I fusion succeeds if and only if the two incoming photons have identical polarization, in which case the two photons are either both transmitted or both reflected such that we retain one photon on each side of the PBS and thus detect exactly one photon. This measures the parity operator $Z_AZ_B$ to be $+1$, a parity measurement which we will use to design purification schemes. More specifically, a successful type-I fusion corresponds to the mapping $\ket{H}\bra{HH}\pm\ket{V}\bra{VV}$, where $\bra{VV}, \bra{HH}$ represent the polarizations of the two photons in the two input arms and $\ket{H},\ket{V}$ represent the photon polarization in the unmeasured output arm. The sign is determined by which of the two detectors clicks.

\subsection{State generation and purification device}
\label{sec_device_scheme}
A single quantum emitter with a spin can generate entangled photonic states deterministically but cannot generate arbitrary entangled states~\cite{Li2022}. For this reason, there are various proposals~\cite{Lee2023, Lobl2024, Gimeno2016, Wein2024, Rimock2026} and experiments~\cite{Meng2023, Thomas2024} that add type-II fusions (Bell-state measurements)~\cite{Browne2005} as a means to generate different states. In Fig.~\ref{fig_setups}(b), we show an exemplary high-level device that can incorporate purification into such generation protocols for entangled photonic states. In this device, one or more quantum emitters in waveguides can be optically excited by a laser, leading to subsequent photon emission. An additional non-resonant laser can be used for optical control of the quantum emitters' spins~\cite{Bodey2019}. The emitted photons are routed by a switching network to (1) type-II fusion setups~\cite{Browne2005} to realize more complex entanglement between spin or photonic qubits~\cite{Gimeno2016, Lobl2024}, (2) type-I fusion setups for purification, (3) internal delay lines that can store photons until a classical control unit decides what to do with them, or (4) output modes releasing the final entangled photonic state~\footnote{The laser light can be routed to the quantum emitters with the same switching network. Alternatively, a separate switching network can be used to reduce laser background affecting the emitted photon.}. The classical control unit collects and processes measurement outcomes and instructs the excitation laser, the spin control laser, and the switching network how to operate on the basis of this information. The shown device allows combining protocols for entangled state generation via quantum emitters and type-II fusions~\cite{Lobl2024, Wein2024} with the purification protocols investigated in this work~\footnote{The graph-state generation protocol from Ref.~\cite{Li2022} could also be realized and combined with type-I based purification if deterministic entangling gates between quantum emitter spins are enabled.}. With the shown device, purification can be performed as the last step or can be applied to intermediate states; several cascaded purifications (see Section~\ref{sec_ghz}) can also be performed. Furthermore, the device can be used to implement several multiplexing strategies to generate purified target states with high probability (see Section~\ref{sec_multiplex}).

\section{Purification of GHZ states}
\label{sec_ghz}
\subsection{The states}
\begin{figure*}[!t]
\includegraphics[width=1.0\textwidth]{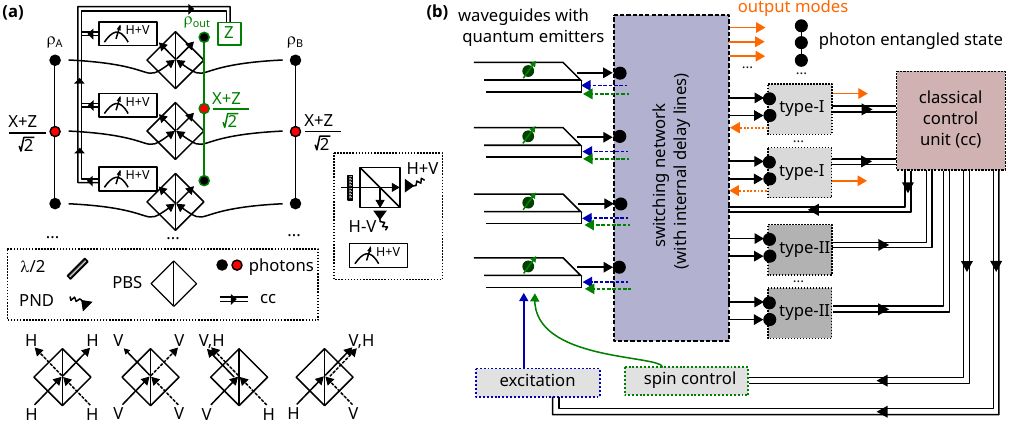}
\caption{\label{fig_setups} Setups for the purification of GHZ or more general CSS states. \textbf{(a)} Purifying the polarization encoded $3$-qubit state $\ket{\Phi_0^+}$, corresponding to a linear graph state with a Hadamard gate $\frac{X+Z}{\sqrt{2}}$ applied to the central qubit. Transversal type-I fusion is implemented by polarizing beam splitters (PBS) and photon-number resolving detectors (PND). The action of a PBS in the $(H, V)$-basis is illustrated on the bottom. The setup for the required $(H+V, H-V)$-basis measurement is shown on the right. When detecting one photon per type-I fusion (heralding criterion), the state in the output mode is kept. The output state is a new copy of $\rho_A/\rho_B$ with reduced undesired contributions from $\ket{\Phi_{i>0}^{\pm}}$. A correction gate conditioned on the measurements outcomes may be required if the state is sent to further cascaded purification setups. For checking the heralding criterion and applying correction gates, the measurement outcomes are processed and forwarded by classical control (cc). By extending the setup with further PBSs in the vertical direction, it can be used for the purification of GHZ states and, more generally, CSS states of arbitrary sizes. \textbf{(b)} High-level illustration of a device that can be used to generate entangled photon states with built-in purification. Photons emitted by the quantum emitters are sent into a switching network (black arrows) and routed to the input modes of type-I/type-II fusions or to output modes (orange arrows). For a type-I fusion, the output mode may be sent back into the switching network for subsequent purification rounds. Photon emission is triggered by resonant excitation lasers, while control of the quantum emitters' spins is achieved by non-resonant lasers. The laser light can be sent to the quantum emitters from the top (solid blue/green arrows) or through the waveguides (dotted blue/green arrows). A classical control unit collects the measured results of the fusions and, on the basis of this information, steers excitation and spin control lasers as well as the switching network.}
\end{figure*}
First, we consider the purification of a state from the class of generalized Bell states over $n$ qubits. These states can be written
\begin{align}
    \ket{\Phi_{k}^{\pm}} = \frac{1}{\sqrt{2}}\left(\ket{k}\pm\ket{2^n-1-k}\right),
\end{align}
where $k\in[0,2^{n-1}-1]$ is an integer and $\ket{k}$ defines the $n$-qubit state according to the binary representation of the number $k$. For $n=3$, $\ket{k=3}$ would, for example, represent the state $\ket{011}$. States with different values of $k$ are orthogonal and all $2^n$ states form a basis of the $n$-qubit Hilbert space. In the described representation $\ket{\Phi_{0}^{+}}=\frac{1}{\sqrt{2}}\left(\ket{0}^{\otimes n}+\ket{1}^{\otimes n}\right)$ is the standard $n$-qubit GHZ state, and for a $3$-qubit system, the different basis states $\ket{\Phi_{k}^{\pm}}$ are
\begin{align}
    &\ket{\Phi_{0}^{\pm}} = \frac{1}{\sqrt{2}}\left(\ket{000}\pm\ket{111}\right)\\
    &\ket{\Phi_{1}^{\pm}} = \frac{1}{\sqrt{2}}\left(\ket{001}\pm\ket{110}\right)\\
    &\ket{\Phi_{2}^{\pm}} = \frac{1}{\sqrt{2}}\left(\ket{010}\pm\ket{101}\right)\\
    &\ket{\Phi_{3}^{\pm}} = \frac{1}{\sqrt{2}}\left(\ket{011}\pm\ket{100}\right).
\end{align}
The computational $0,1$ states of the qubits can e.g. correspond to the photon polarization ($H, V$), the photon modes in path encoding, or the early and late ($e, l$) time bins in time bin encoding.

We consider here a protocol for purifying the $n$-qubit GHZ state $\ket{\Phi_{0}^{+}}$, but other states $\ket{\Phi_{i}^{\pm}}$ can be purified analogously. We assume that the desired states are imperfect, that is, they have fidelity $F<1.0$. We also assume that the corresponding mixed state has a density matrix that is diagonal in the generalized Bell-state basis and thus can be expressed as
\begin{align}
\label{eq_rho}
&\rho = F\ket{\Phi_{0}^+}\bra{\Phi_{0}^+} + \alpha_{0,-}\ket{\Phi_0^-}\bra{\Phi_0^-}\notag\\
&+\sum_{k=1}^{2^{n-1}-1}\left(\alpha_{k,+}\ket{\Phi_k^+}\bra{\Phi_k^+}+\alpha_{k,-}\ket{\Phi_k^-}\bra{\Phi_k^-}\right),
\end{align}
with $\alpha_{0,-} + \sum_{k=1}^{2^{n-1}-1}\left(\alpha_{k,+}+\alpha_{k,-}\right) = 1-F$. The diagonal density matrix in Eq.~\eqref{eq_rho} would arise for uncorrelated bit- and phase-flip errors. Note that if it had nonzero off-diagonal elements, it can be made diagonal by applying the stabilizer generators of $\ket{\Phi_0^+}$ probabilistically~\cite{Aschauer2005, Bennett1996}. (The stabilizer $\mathcal{S}_k$ of $\ket{\Phi_{k}^{\pm}}$ is the group of operators $S_k \in\mathcal{S}_k$ with $S_k\ket{\Phi_{k}^{\pm}} = \ket{\Phi_{k}^{\pm}}$.~\footnote{Here, the stabilizer generators have the form $\mathcal{S}_k =\langle\pm X^{\otimes n},(-1)^{k_i+k_{i+1}}Z_iZ_{i+1}\rangle$ 
with $i\in[1,n)$ and $k_i$ being the i\textit{th} element in the bitwise representation of $k$.}). The fidelity calculated here thus represents a lower bound on the achievable fidelity for a density matrix with diagonal elements $\alpha_{k,\pm}$.

Next, we will describe how the state can be purified, which is performed in two rounds. In the first round, bit-flip errors are reduced by suppressing terms involving $\ket{\Phi_{k>0}^{\pm}}$. The second purification round reduces phase-flip errors, filtering out the undesired state $\ket{\Phi_{0}^{-}}$. Depending on the main source of infidelity, one purification round can be left out, or the two rounds can be performed in opposite order. To improve fidelities, several purification rounds can be concatenated/cascaded.

\subsection{Reducing bit-flip errors}
Assume that there are two copies $\rho_A, \rho_B$ of the state in Eq.~\eqref{eq_rho}. For polarization encoded qubits, the photons of the two states are sent into one of the transversal type-I fusion setups shown in Fig.\ref{fig_setups}(a,b). The setup is shown for three-qubit input states, but can be extended to any number of qubits by repeating type-I fusion setups in the vertical direction.

To understand the effect of transversal type-I fusion on the input state, we consider several cross-terms from $\rho_A\otimes \rho_B$ independently. For all summands of the cross term $\ket{\Phi_i^{\pm}}_A\otimes\ket{\Phi_j^{\pm}}_B$ with $i\neq j$, at least one qubit in $A$ differs in polarization from the corresponding qubit in $B$. As illustrated at the bottom of Fig.~\ref{fig_setups}(a), the PBSs have the important property that they lead to two photons in different output ports if and only if the polarization of the two input modes coincides. Therefore, at least one detector in Fig.~\ref{fig_setups}(a) will detect zero or two photons for $i\neq j$. This is what enables filtering out cross terms where one term has a bit-flip error. In contrast, the term $\ket{\Phi_i^{\pm}}_A\otimes\ket{\Phi_i^{\pm}}_B$ will lead to exactly one photon per detector in part $A$ while outputting a new state $\ket{\Phi_i^{\pm}}_B$ (up to heralded local gates).

In particular, sending the state $\ket{\Phi_0^+}_A\otimes\ket{\Phi_0^+}_B = \frac{1}{\sqrt{2}}\left(\ket{HH..H}+\ket{VV..V}\right)_A \otimes \frac{1}{\sqrt{2}}\left(\ket{HH..H}+\ket{VV..V}\right)_B$ to the polarizing beam splitters leads to the state
\begin{align}
    &\frac{1}{2}\ket{HH..H}_A\ket{HH..H}_B\label{eq_after_pbs_1}\\
    +&\frac{1}{2}\ket{VV..V}_A\ket{VV..V}_B\label{eq_after_pbs_2}\\
    +&\frac{1}{2}\ket{HH..H}_A\ket{VV..V}_A\label{eq_after_pbs_3}\\
    +&\frac{1}{2}\ket{HH..H}_B\ket{VV..V}_B\label{eq_after_pbs_4}.
\end{align}
The photons in part $A$ are then measured in the $(H+V, H-V)$-basis. With probability $0.5$, one obtains a measurement outcome corresponding to the terms in Eq.~\eqref{eq_after_pbs_1},~\eqref{eq_after_pbs_2}. In this case, the measurement is successful, meaning that exactly one photon is measured per detector. The resulting state is $\frac{1}{\sqrt{2}}\left(\ket{HH..H}\pm\ket{VV..V}\right)_B$ depending on whether an even ($+$) or odd ($-$) number of photons are projectively measured to be $H-V$. Up to a heralded local $Z$-gate, the resulting state is thus $\ket{\Phi_0^+}$. If the output state is sent to further purification setups, this gate must be corrected conditioned on the measurement outcomes as illustrated in Fig.~\ref{fig_setups}. In the failure case (probability $0.5$), the measurement projects onto one of the terms from Eqs.~\eqref{eq_after_pbs_3},~\eqref{eq_after_pbs_4}. In this case, either all photons are destructively measured, or the output state is not entangled. The purification attempt is then discarded.

\begin{figure}
\includegraphics[width=1.0\linewidth]{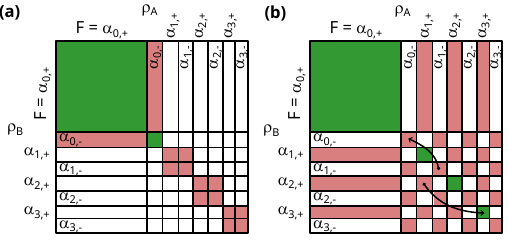}
\caption{\label{fig_illustration}\textbf{(a)} Illustration of what happens to the various cross-terms in one round of bit-flip purification of a 3-photon GHZ state. Terms highlighted in red lead to an undesired contribution in the final state, terms highlighted in green lead to the desired state, and all other terms are filtered out as the detection patterns do not meet the heralding criterion. \textbf{(b)} Same illustration for one round of phase-flip purification. Here, cross terms of the form $\ket{\Phi_{i>0}^+}_A\ket{\Phi_{i>0}^+}_B$ also lead to the desired state $\ket{\Phi_{0}^+}$. More generally, cross terms of the form $\ket{\Phi_i^{\pm}}_A\ket{\Phi_j^{\pm}}_B$ lead to the state $\ket{\Phi_{i\oplus j}^{\pm}}$, as exemplary illustrated by the two arrows.}
\end{figure}

More generally, cross-terms of the form $\ket{\Phi_i^{\pm}}_A\otimes\ket{\Phi_i^{\pm}}_B$ lead to the state $\ket{\Phi_i^{+}}_B$, passing the purification protocol without heralding an error. This leads to an error for $i\neq 0$ as illustrated in Fig.~\ref{fig_illustration}(a) on the diagonal. However, the weight of these errors in the new state is determined by $\alpha_{i,\pm}^2$ and therefore is suppressed by the purification for $F\rightarrow 1$, i.e. $\alpha_{i,\pm} \ll 1$. Furthermore, cross-terms of the form $\ket{\Phi_i^{\pm}}_A\otimes\ket{\Phi_i^{\mp}}_B$ also lead to one photon per detector with a probability of $0.5$. Therefore, the cross terms $\ket{\Phi_i^+}_A\otimes\ket{\Phi_i^{-}}_B$ and $\ket{\Phi_i^-}_A\otimes\ket{\Phi_i^{+}}_B$ cannot be filtered out by selecting states based on the heralding criterion. When part $A$ is measured and the corresponding local correction gates are applied, this leads to an additional term $\ket{\Phi_i^-}_B$ (minus sign because the corresponding cross-terms have a prefactor $-1$ in Eq.~\eqref{eq_after_pbs_2}) with probability $2F\alpha_i^-$. The phase-flip term $\ket{\Phi_i^-}_B$ is therefore not suppressed by the purification for $F\rightarrow 1$.

In summary, the first purification round reduces bit-flip errors, converting the state $\rho_A\otimes\rho_B$ in Eq.~\eqref{eq_rho} into
\begin{align}\label{eq_purify1}
    &\rho_B = \frac{F^2+\alpha_{0,-}^2}{\sigma}\ket{\Phi_{0}^+}\bra{\Phi_{0}^+} + \frac{2F\alpha_{0,-}}{\sigma}\ket{\Phi_{0}^-}\bra{\Phi_{0}^-}\notag\\
&+\frac{1}{\sigma}\sum_{k=1}^{2^{n-1}-1}\left((\alpha_{k,+}^2+\alpha_{k,-}^2)\ket{\Phi_k^+}\bra{\Phi_k^+}\right)\notag\\
&+\frac{1}{\sigma}\sum_{k=1}^{2^{n-1}-1}\left(2\alpha_{k,+}\alpha_{k,-}\ket{\Phi_k^-}\bra{\Phi_k^-}\right),
\end{align}
with probability $\sigma/2$, where $\sigma=F^2+2F\alpha_{0,-}+\alpha_{0,-}^2+\sum_{k=1}^{2^{n-1}-1}\left(\alpha_{k,+}^2+\alpha_{k,-}^2+2\alpha_{k,+}\alpha_{k,-}\right)$. In Fig.~\ref{fig_illustration}(a) we illustrate for the various cross-terms whether they lead to the desired state, an unintended state, or whether they are filtered out by the heralding criterion not being met (post-selection~\footnote{All our schemes are heralded, meaning that, when a measurement matches a heralding criterion, an unmeasured/alive photonic output state is generated. In agreement with corresponding literature on error correction~\cite{Goyal2006}, we say that the state is post-selected (upon the heralding criterion). This is not to be confused with post-selecting on destructive measurement outcomes applied to the target state itself~\cite{Pan2003}.}). In the absence of phase-flip errors ($\alpha_{0,-}=0$), one can easily see from Eq.~\eqref{eq_purify1} that, after purification, the new fidelity $F^2/\sigma$ is guaranteed to be higher than the initial fidelity $F$ if $F>0.5$~\cite{Pan2001}. However, if $\alpha_{0,-}>0$, phase-flip errors are amplified by reducing bit-flip errors (cf. the term $2F\alpha_{0,-}$ in Eq.~\eqref{eq_purify1}). For a small identical rate of Pauli $X$- and $Z$-errors, for instance, we find cases where a single purification round does only very slightly improve fidelity or even makes it worse. To reliably improve fidelities over a broad parameter regime, a bit-flip purification round, therefore, has to be combined with an additional phase-flip purification round.

\subsection{Reducing phase-flip errors}
\label{sec_phase_ghz}
The purification discussed in the previous section cannot correct phase-flip errors, i.e. cross-terms of the form $\ket{\Phi_i^{\pm}}_A\ket{\Phi_i^{\mp}}_B$ are not filtered out by post-selecting a state upon the right detection pattern (heralding criterion). Instead, the purification of phase-flip errors can be performed by applying a transversal Hadamard gate to all qubits in $\rho_A, \rho_B$ and then sending the photons into the setup in Fig.~\ref{fig_setups}(a). A single-qubit Hadamard gate has the form $H=\frac{1}{\sqrt{2}}(X+Z)$, where $X, Z$ are Pauli matrices. A transversal Hadamard gate, i.e. a single-qubit Hadamard gate on every individual qubit, converts phase-flip into bit flip errors, which is why subsequent bit-flip purification effectively corrects for phase-flips. For polarization encoding, the required Hadamard gates can be realized by half-waveplates. In the following, we analyze this phase-flip purification scheme for GHZ states.

We consider two input states given by Eq.~\eqref{eq_rho}, and consider the various cross-terms of the corresponding tensor product. First, we note the following relation:
\begin{align}
    H^{\otimes n}\ket{\Phi_0^{\pm}}=\frac{1}{\sqrt{2^{n+1}}}\left(\sum_{k=0}^{2^n-1}\ket{k}\pm \sum_{k=0}^{2^n-1}(-1)^{q_k}\ket{k}\right),
\end{align}
where $q_k$ is the digit sum of the binary representation of the number $k$. Written differently, we have:
\begin{align}
    H^{\otimes n}\ket{\Phi_0^+}=\frac{1}{\sqrt{2^{n-1}}} \sum_{k=0}^{2^n-1}E(q_k)\ket{k}\label{eq_even}\\
    H^{\otimes n}\ket{\Phi_0^-}=\frac{1}{\sqrt{2^{n-1}}} \sum_{k=0}^{2^n-1}(1-E(q_k))\ket{k}\label{eq_odd},
\end{align}
where $E(x)=1$ if $x$ is even and $E(x)=0$ when $x$ is odd. Thus, $H^{\otimes n}\ket{\Phi_0^+}$ corresponds to a sum over all state vectors with an even digit sum, while $H^{\otimes n}\ket{\Phi_0^-}$ is a sum over all state vectors with odd digit sum. Therefore, cross terms of the form $H^{\otimes n}\ket{\Phi_0^{\pm}}_A\otimes H^{\otimes n}\ket{\Phi_0^{\mp}}_B$ differ in at least one qubit between $A, B$, leading to at least one detector on which zero or two photons will be detected. This reduces phase-flip errors by filtering out these cross-terms. 

For more general cross-terms, we use the relation $H^{\otimes n}\ket{\Phi_j^{\pm}} = H^{\otimes n}X^{\vec{j}}\ket{\Phi_0^{\pm}} = Z^{\vec{j}}H^{\otimes n}\ket{\Phi_0^{\pm}}$, where $X^{\vec{j}}$ represents a product of Pauli $X$ operators applied to all qubits for which the binary representation of $j$ is nonzero. With this relation, we can generalize Eq.~\eqref{eq_odd} and Eq.~\eqref{eq_even} to
\begin{align}
    H^{\otimes n}\ket{\Phi_j^+}&=\frac{Z^{\vec{j}}}{\sqrt{2^{n-1}}} \sum_{k=0}^{2^n-1}E(q_k)\ket{k}\notag\\
    &=\frac{1}{\sqrt{2^{n-1}}} \sum_{k=0}^{2^n-1}(-1)^{q_{k\land j}}E(q_k)\ket{k}\label{eq_even_general}\\
    H^{\otimes n}\ket{\Phi_j^-}&=\frac{1}{\sqrt{2^{n-1}}} \sum_{k=0}^{2^n-1}(-1)^{q_{k\land j}}(1-E(q_k))\ket{k}\label{eq_odd_general},
\end{align}
where $k\land j$ represents the bitwise AND operation applied to the binary representations of $k,j$.

These relations show that all cross terms of the form $H^{\otimes n}\ket{\Phi_i^{\pm}}_A\otimes H^{\otimes n}\ket{\Phi_j^{\mp}}_B$ are filtered out since Eq.~\eqref{eq_even_general} and Eq.~\eqref{eq_odd_general} do not share any identical basis states for which the setups in Fig.~\ref{fig_setups} would detect one event per detector.

Now we consider cross-terms that are not filtered out. As an example, first consider the term $H^{\otimes n}\ket{\Phi_0^+}_A\otimes H^{\otimes n}\ket{\Phi_0^+}_B$. With probability $1/2^{n-1}$, two summands from Eq.~\eqref{eq_even} have identical terms for $\ket{k}_A$ and $\ket{k}_B$ in which case there is one click for every detector. The resulting state is $Z^{\vec{m}}H^{\otimes n}\ket{\Phi_0^+}_B$, where $\vec{m} \in \{0,1\}^n$ are the measurement results in part $A$ with $0$ representing $\ket{+}=\frac{1}{\sqrt{2}}(\ket{H}+\ket{V})$ and $1$ representing $\ket{-}=\frac{1}{\sqrt{2}}(\ket{H}-\ket{V})$ being measured. The same argument applies to the term $H^{\otimes n}\ket{\Phi_0^-}_A\otimes H^{\otimes n}\ket{\Phi_0^-}_B$ for which one obtains $Z^{\vec{m}}H^{\otimes n}\ket{\Phi_0^-}_B$. Note that $H^{\otimes n}\ket{\Phi_{i>0}^{\pm}}$ differs from $H^{\otimes n}\ket{\Phi_0^{\pm}}$ only by some summands having a different sign. These signs cancel for the terms in $H^{\otimes n}\ket{\Phi_{i>0}^{\pm}}_A\otimes H^{\otimes n}\ket{\Phi_{i>0}^{\pm}}_B$ that have identical $\ket{k_A}, \ket{k_B}$. Therefore, the terms $H^{\otimes n}\ket{\Phi_{i>0}^{\pm}}_A\otimes H^{\otimes n}\ket{\Phi_{i>0}^{\pm}}_B$ result in $Z^{\vec{m}}H^{\otimes n}\ket{\Phi_0^{\pm}}_B$ with probability $1/2^{n-1}$.

Considering Eq.~\eqref{eq_even_general} and Eq.~\eqref{eq_odd_general}, one can see that, more generally, the cross-terms $H^{\otimes n}\ket{\Phi_{i}^{\pm}}_A\otimes H^{\otimes n}\ket{\Phi_j^{\pm}}_B$ lead, with probability $1/2^{n-1}$, to all detectors detecting one photon (when the basis states $\ket{k}$ of both terms coincide). Using the relation $(-1)^{q_{k\land j}}\cdot(-1)^{q_{k\land i}} = (-1)^{q_{k\land (i\oplus j)}}$, we find that the obtained state is $Z^{\vec{m}}H^{\otimes n}\ket{\Phi_{i\oplus j}^{\pm}}_B$, where $i\oplus j$ represents the bitwise XOR, i.e. addition without carryover. Starting from Eq.~\eqref{eq_rho} where $\alpha_{0,+}\equiv F$ is the fidelity of the initial state, the final state $\rho_B$ is thus given by
\begin{align}
\label{eq_purify_2}
    \begin{split}
    &H^{\otimes n}\rho_B H^{\otimes n}=\frac{1}{\sigma}\sum_{k=0}^{2^{n-1}-1}\sum_{j=0}^{2^{n-1}-1} \\
    &\left(\alpha_{k,+}\alpha_{j,+}\ket{\Phi_{k\oplus j}^+}\bra{\Phi_{k\oplus j}^+} +
    \alpha_{k,-}\alpha_{j,-}\ket{\Phi_{k\oplus j}^-}\bra{\Phi_{k\oplus j}^-}\right).
    \end{split}
\end{align}
This final state is successfully obtained with probability $\sigma/2^{n-1}$, where $\sigma=\sum_{k=0}^{2^{n-1}-1}\sum_{j=0}^{2^{n-1}-1}\left(\alpha_{k,+}\alpha_{j,+} +\alpha_{k,-}\alpha_{j,-}\right)$. The new fidelity of the desired state $\ket{\Phi_{0}^+}$ thus becomes $\frac{1}{\sigma}\left(F^2+\sum_{k=1}^{2^{n-1}-1}\alpha_{k,+}^2\right)$, which is above the original fidelity if phase-flip errors dominate. However, the phase-flip-error purification scheme amplifies bit-flip errors, and in Fig.~\ref{fig_illustration}(b), we illustrate which of the cross-terms leads to the correct state, an unintended state, and which are filtered out. In Section~\ref{sec_numerics}, we will investigate numerically where alternating rounds of bit- and phase-flip-correcting purification rounds lead to improved fidelity.

\section{Graph and CSS states}
\label{sec_graph_css_def}
Using type-II fusions~\cite{Browne2005, Gimeno2016, Lobl2024c}, one can build any graph state from ($n\geq3$)-GHZ states~\cite{Lee2023,Lobl2023}. In principle, purification of GHZ states is thus sufficient. However, direct purification protocols for more general states are desirable because building states from small GHZ states can require many type-II fusions~\cite{Lee2023}, leading to error accumulation. In the following two sections, we will discuss the purification of more general states, in particular, graph states and CSS states, which we introduce below.

Given an undirected graph $G=(V,E)$ with $V$ being its vertices and $E$ its edges, a graph state $\ket{G}$ is defined as $1/\sqrt{2^n}\prod_{(i,j)\in E}CZ_{i,j}(\ket{0}+\ket{1})^{\otimes \mid V \mid}$, where $CZ_{i,j}$ is a controlled $Z$-gate~\cite{Hein2004}. Graph states are eigenstates of the operators (stabilizer generators) $X_i\prod_{(i,j)\in E}Z_j$~\cite{Looi2008} with eigenvalues $+1$ and, therefore, they are stabilizer states~\cite{Aaronson2004}. Furthermore, every stabilizer state is locally equivalent to a graph state~\cite{Nest2004}.

In analogy to a Calderbank–Shor–Steane (CSS) code~\cite{Calderbank1996, Steane1996}, we define a CSS state as a stabilizer state that has a set of stabilizer generators that are products of only Pauli $X$ or only Pauli $Z$ operators. One can easily see that every two-colorable graph state is locally equivalent to a CSS state by applying Hadamard gates on all qubits of one color. It turns out that this also applies in the opposite direction, so, every CSS state is equivalent up to local Clifford gates to a two-colorable graph state and vice versa~\cite{Chen2004}. Therefore, an equivalence class of stabilizer states contains a CSS state if and only if it contains a two-colorable graph state. Thus, a purification protocol for a CSS state also enables purification of any locally equivalent two-colorable graph state by first purifying the CSS states and then converting it into the two-colorable graph state by local gates.

All stabilizer states are locally equivalent to graph states~\cite{Nest2004}, but not all are locally equivalent to CSS states (or two-colorable graph states). Therefore, we investigate separately the purification of CSS states (Section~\ref{sec_css}) and the purification of arbitrary graph states (Section~\ref{sec_all_graph}).

\section{Purifying CSS states}
\label{sec_css}
Here, we consider the purification of an $n$-qubit CSS state. Most multipartite purification protocols have been analyzed in the picture of two-colorable graph states~\cite{Dur2003, Aschauer2005, Goyal2006}. Due to the local Clifford equivalence between the two classes of states~\cite{Chen2004}, the corresponding purification protocols can be derived analogously~\cite{Dur2003, Aschauer2005}. We consider CSS states because they are the most convenient choice when using type-I fusions. We will first discuss the purification schemes using deterministic CNOT gates~\cite{Dur2003, Aschauer2005} and then show how they can be realized without deterministic entangling gates, i.e. via type-I fusions.

\subsection{Purification with deterministic CNOT gates}
We define the CSS state $\ket{C_{a_x, a_z}}$ as the stabilizer state with $n_x$ independent $X$-type stabilizer generators $\mathcal{X}=\{X_{S_1}...X_{S_{n_x}}\}$ and $n_z=n-n_x$ independent $Z$-type stabilizer generators $\mathcal{Z}=\{Z_{Q_1}...Z_{Q_{n_z}}\}$, where $X_{S_i}=(-1)^{a_x[i]}\prod_{j\in S_i} X_j$ and $Z_{Q_i}=(-1)^{a_z[i]}\prod_{j\in Q_i} Z_j$ with $a_x \in \{0,1\}^{n_x}$ and $a_z \in \{0,1\}^{n_z}$. Bit-flip (Pauli-X) errors flip the sign of some stabilizers in $\mathcal{Z}$, i.e., apply a logical NOT to some of the elements of $a_z$. Phase-flip (Pauli-Z) errors flip the sign of some stabilizers in $\mathcal{X}$, i.e. apply a logical NOT to some of the elements of $a_x$. All $2^n$ states of type $\ket{C_{a_x, a_z}}$ form a basis of the $n$-qubit Hilbert space. Therefore, we can consider purification of a mixed state like the one in Eq.~\ref{eq_rho} where the generalized Bell states are replaced by the states $\ket{C_{a_x, a_z}}$. We thus assume two copies of a mixed state, here diagonal in the CSS state basis:
\begin{equation}
\label{eq_rho_css}
\rho=\sum_{a_x,a_z}\alpha_{a_x,a_z}\ket{C_{a_x,a_z}}\bra{C_{a_x,a_z}},
\end{equation}
where $\sum \alpha_{a_x,a_z} = 1$ and thus the fidelity of the desired state $\ket{C_{a_x,a_z}}$ is given by $F=\alpha_{0,0}$.

In the following, we will use the symplectic representation for stabilizer states, which represents an $n$-qubit stabilizer state as a $n\times 2n$ matrix over $\mathbb{F}_2$ and a $n$-vector over $\mathbb{F}_2$~\cite{Aaronson2004}. For the CSS state $\ket{C_{a_x, a_z}}$, the symplectic representation is
\begin{equation}
    \begin{bmatrix}
    a_x\\
    a_z
    \end{bmatrix}\in \mathbb{F}_2^n,
    \begin{bmatrix}[c|c]
    H_x & 0\\
    0 & H_z
    \end{bmatrix}\in \mathbb{F}_2^{n\times 2n},
\end{equation}
where $H_x$ is an $n_x\times n$ and $H_z$ an $n_z\times n$ matrix. In this representation, every row of the matrix is a stabilizer with the first $n$ columns representing Pauli $X$-gates and the second $n$ columns representing Pauli $Z$ gates. The $\mathbb{F}_2^n$ vector represents the corresponding stabilizer signs.

Cross-terms of the form $\ket{C_{a_x, a_z}}\otimes \ket{C_{b_x, b_z}}$ thus have the symplectic representation
\begin{equation}
    \label{eq_css_start}
    \begin{bmatrix}
    a_x\\
    b_x\\
    a_z\\
    b_z
    \end{bmatrix},
    \begin{bmatrix}[cc|cc]
    H_x & 0 & 0 & 0\\
    0 & H_x & 0 & 0\\
    0 & 0 & H_z & 0\\
    0 & 0 & 0 & H_z
    \end{bmatrix}.
\end{equation}
We apply transversal CNOT gates with the first $n$ qubits from $\ket{C_{a_x, a_z}}$ as the control and the qubits from $\ket{C_{b_x, b_z}}$ as the target. Note that a single gate $\text{CNOT}_{a,b}$ corresponds to the following operation on the symplectic matrix~\cite{Aaronson2004}: $r_i\rightarrow r_i \oplus x_{ia}z_{ib}(x_{ib}\oplus z_{ia}\oplus 1)$, $x_{ib}\rightarrow x_{ib}\oplus x_{ia}$, $z_{ia}\rightarrow z_{ia}\oplus z_{ib}$. Therefore, the transversal CNOTs lead to the state
\begin{equation}
    \begin{bmatrix}
    a_x\\
    b_x\\
    a_z\\
    b_z
    \end{bmatrix},
    \begin{bmatrix}[cc|cc]
    H_x & H_x & 0 & 0\\
    0 & H_x & 0 & 0\\
    0 & 0 & H_z & 0\\
    0 & 0 & H_z & H_z
    \end{bmatrix}.
\end{equation}
By multiplying, i.e. modulo-two addition in the symplectic representation, the stabilizers in the second(third) row on those in the first(fourth) row, one finds that the obtained state is identical to
\begin{equation}
    \begin{bmatrix}
    a_x\oplus b_x\\
    b_x\\
    a_z\\
    b_z\oplus a_z
    \end{bmatrix},
    \begin{bmatrix}[cc|cc]
    H_x & 0 & 0 & 0\\
    0 & H_x & 0 & 0\\
    0 & 0 & H_z & 0\\
    0 & 0 & 0 & H_z
    \end{bmatrix}.
\end{equation}
This stabilizer tableau corresponds to the state~\cite{Dur2003, Aschauer2005}~\footnote{The result also shows that a logical CNOT gate between two CSS codes can be realized by transversal CNOT gates between them~\cite{Gottesman1997}.}
\begin{equation}
    \label{eq_binary_add}
    \ket{C_{a_x\oplus b_x, a_z}}\otimes \ket{C_{b_x, b_z\oplus a_z}},
\end{equation}
where both states are again unentangled~\footnote{In the analogous protocol for two-colorable graph states~\cite{Dur2003, Aschauer2005}, it can also be seen in a graph state picture that both systems are unentangled after applying all CNOT gates, in particular, by using Eq. (7) from Ref.~\cite{Doherty2026}.}, but information about stabilizer sign has been transferred. Now, one can measure all qubits of the second subsystem in the Pauli-$Z$ basis and postselect on the measurement outcomes where the correspondingly measured value of the $Z$-type stabilizers is $+1$. This is a post-selection  of cross-terms with $a_z\oplus b_z=0$, thus reducing bit-flip errors on all qubits~\footnote{We can assume that every qubit has support on at least one $Z$-type ($X$-type) stabilizer generator, and therefore any single-qubit bit-flip (phase-flip) error is heralded. If a qubit $k$ had only support on $X$-type ($Z$-type) stabilizers, one would find by Gaussian elimination a set of generators where it only has support on one stabilizer. The other stabilizers would have full rank on the remaining qubits and $k$ would thus be an isolated/unentangled qubit.}. For the cross-term considered, one obtains the output state $\prod_j m_j\cdot\ket{C_{a_x\oplus b_x, a_z}}$, with $m_j\in\{-1,+1\}$ being the individual measurement results that define a global phase factor. Instead, one can measure all qubits of the first subsystem in the Pauli-$X$ basis and postselect on the measurement outcomes where the correspondingly measured value of the $X$-type stabilizers is $+1$. This is a post-selection  of cross-terms with $a_x\oplus b_x=0$, thus reducing phase-flip errors. The output state is $\prod_j m_j\cdot\ket{C_{b_x, b_z\oplus a_z}}$, with $m_j\in\{-1,+1\}$ being the individual measurement results on the $j$th qubits.

Knowing the effect of the purification circuits on pure states, we can evaluate their effect on the mixed state in Eq.~\eqref{eq_rho_css}. Applying bit-flip purification leads to the following new coefficients of the mixed state~\cite{Dur2003, Aschauer2005} 
\begin{equation}
    \alpha_{a_x',a_z'}=\frac{1}{\sigma_b}\sum_{a_x,b_x: (a_x\oplus b_x)=a_x'}\alpha_{a_x,a_z'}\cdot\alpha_{b_x,a_z'},
\end{equation}
with $\sigma_b$ normalizing the probability distribution of the new mixed state. The value of $\sigma_b$ corresponds to the probability that the bit-flip purification is successful without error detection. Instead, applying phase-flip purification gives
\begin{equation}
    \alpha_{a_x',a_z'}=\frac{1}{\sigma_p}\sum_{a_z,b_z: (a_z\oplus b_z)=a_z'}\alpha_{a_x',a_z}\cdot\alpha_{a_x',b_z},
\end{equation}
where the role of $X$- and $Z$-type stabilizers is interchanged by the initial transversal Hadamard gates. The normalization factor $\sigma_p$ corresponds to the probability that phase-flip purification succeeds without error detection.

\subsection{Implementation with type-I fusions}
\label{sec_type1_scheme}
The described purification scheme uses transversal CNOT gates and Pauli measurements of one of the two subsystems~\cite{Dur2003, Aschauer2005}. Although technically possible~\cite{Hacker2016}, CNOT gates are very challenging to implement for photonic qubits. However, an equivalent circuit using type-I fusions can be found by the following observation. 

Assume we apply  $\text{CNOT}_{1,2}=\ket{00}\bra{00}+\ket{01}\bra{01}+\ket{11}\bra{10}+\ket{10}\bra{11}$, then apply the gate $Z_1^m$ ($1_1$ if $m=0$, $Z_1$ if $m=1$), and finally measure the second qubit in the $Z$-basis. When the second qubit is measured in $\ket{0}$, this corresponds to the operation $\ket{0}\bra{00}+(-1)^m\ket{1}\bra{11}$. When the second qubit is measured in $\ket{1}$, the operation is $\ket{0}\bra{01}+(-1)^m\ket{1}\bra{10}$. For implementing the operation with successful type-I fusion, we can focus on the first case because input states $\ket{01}=\ket{HV}, \ket{10}=\ket{VH}$ lead to failure of type-I fusion. Successful type-I fusion implements the map
\begin{equation}
\ket{H}\bra{HH}+(-1)^m\ket{V}\bra{VV},
\end{equation}
where $m$ specifies the result of the projective measurement of one photon in type-I fusion ($m=0$ for $\ket{+}$, $m=1$ for $\ket{-}$). Defining $\ket{0}=\ket{H}, \ket{1}=\ket{V}$, this map coincides with the above operation, i.e. $\text{CNOT}_{1,2}$ followed by the gate $Z_1^m$, and measuring qubit $2$ to be $\ket{0}_2$~\footnote{The above relation can also be seen by noting that CNOT and $Z$-measurement implement a measurement of the parity $Z_AZ_B$ which is heralded by the success of type-I fusion (see Section~\ref{sec_type1}).}. When type-I fusion measures the photon in $\ket{-}$ ($m=1$) the measurement-dependent correction gate $Z_1$ must be applied to the first qubit to obtain a deterministic state. In our application, a sequence of successful transversal type-I fusions thus implements, up to heralded Pauli correction gates, the same operation as the transversal CNOT gates followed by $Z$-measurements on the second subsystem.

This implies that the transversal type-I fusion from Fig.~\ref{fig_setups} with corresponding postselection reduces bit-flip errors. In the previous section, we have also described a method for reducing phase-flip errors that uses transversal CNOT gates followed by $X$-basis measurements on the first subsystem. To keep the protocol compatible with type-I fusions, we instead assume that transversal Hadamard operations are applied before the type-I fusion, converting bit-flip errors to phase-flip errors. Keeping the rest of the protocol identical, this gives a protocol for correcting phase-flip errors.

Type-I fusions only succeed probabilistically, and thus we need to clarify the overall success probabilities of the general CSS purification schemes. The following explanation considers the effect of type-I fusions on the stabilizer tableau, but the success probabilities can also be understood in the Schr{\"o}dinger picture (see Appendix~\ref{CSS_probabiliy}). First, we note that type-I succeeds (fails) deterministically if and only if $+Z_AZ_B$ ($-Z_AZ_B$) is a stabilizer of the state. Second, for stabilizer states, type-I fusion is deterministic or has a success probability of $p_s=1/2$. This is because the involved single-qubit measurement can only have the probabilities $0,0.5,1$ to measure the state $\ket{0}$ (which follows from Theorem 8 in Ref.~\cite{Garcia2012}). Starting from the parity check matrix with $2n$ rows in Eq.~\ref{eq_css_start}, we apply the $n$ type-I fusions one by one, resulting in a parity check matrix with $n$ rows. Since one qubit per fusion is measured destructively, one row and two columns are removed in every step. Since we finally obtain the same state as the two copies, $n_x$ rows must be removed from the top left ($X$-type) and $n_z=n-n_x$ rows must be removed from the bottom right ($Z$-type) part of the parity check matrix. A successful type-I fusion between qubits $a_i, b_i$ corresponds to $\text{CNOT}_{a_i, b_i}$ and a $Z$-basis measurement on $b_i$ where the measured stabilizer is not added to the tableau as the photon measurement is destructive. Whenever there is at least one stabilizer in the $X$-part with support on $X_{a_i}\mathds{1}_{b_i}$ or $\mathds{1}_{a_i}X_{b_i}$, the CNOT plus $Z$-measurement (type-I fusion) will lead to the removal of one stabilizer from the $X$-part (see Ref.~\cite{Aaronson2004}). At the same time, $\pm Z_{a_i}Z_{b_i}$ cannot be a stabilizer as it anticommutes with $X_{a_i}\mathds{1}_{b_i}$ and $\mathds{1}_{a_i}X_{b_i}$. The type-I fusion will therefore be probabilistic for the $n_x$ fusion events in which the $X$-type block of the stabilizer tableau shrinks. The fusion succeeds with $1/2$ in all these cases.

If there is no stabilizer with support on $X_{a_i}\mathds{1}_{b_i}$ or $\mathds{1}_{a_i}X_{b_i}$, the type-I fusion (CNOT with $Z$-measurement) does not shrink the $X$-type part of the stabilizer tableau. Since one qubit is removed, one row must be removed from the $Z$-type part of the tableau in that case. This happens if and only if $\pm Z_{a_i}Z_{b_i}$ is a stabilizer and therefore exactly in the cases where the type-I fusion is deterministic. Therefore, the probability that all type-I fusions are successful is $1/2^{n_x}$ (unless it is zero due to deterministic failure of one fusion). The overall success probability for the bit-flip correction is the product of this value with the probability $\sigma_b$ that no bit-flip error is detected:
\begin{equation}
    p_s^{(b)}=\sigma_b/2^{n_x}.
\end{equation}
With the same argument, one obtains
\begin{equation}
    \label{eq:phase}
    p_s^{(p)}=\sigma_p/2^{n_z}
\end{equation}
for the success probability of the phase-flip correction, with $\sigma_p$ being the probability that no phase-flip error is detected. These results are consistent with the purification scheme for GHZ-states in Section~\ref{sec_ghz} which are CSS stabilizer states with $n_x=1$ and $n_z=n-1$.

\subsection{Generalized DEJMPS scheme}
\label{sec_dejmps}
For Bell states, a different purification scheme has been given in Ref.~\cite{Deutsch1996}. It has been shown that this scheme, which we will refer to as the DEJMPS scheme, can improve the convergence towards unit fidelity when Bell states are purified in cascaded purification rounds~\cite{Dur1999}. We present a generalization of this scheme that applies to a specific class of CSS states. We consider two CSS states as in Eq.~\eqref{eq_css_start} but assume that the $X$- and $Z$-type stabilizer generators exactly coincide ($H_x=H_z$)~\footnote{The condition is met for Bell states, because the $X$- and the $Z$-type stabilizer have support on both qubits.}. Applying the gate $U=e^{i\frac{\pi}{4}X}$ transversally to both states leads to the state
\begin{equation}
    \begin{bmatrix}
    a_x\\
    b_x\\
    a_z\\
    b_z
    \end{bmatrix},
    \begin{bmatrix}[cc|cc]
    H_x & 0 & 0 & 0\\
    0 & H_x & 0 & 0\\
    H_x & 0 & H_x & 0\\
    0 & H_x & 0 & H_x
    \end{bmatrix}.
\end{equation}
Here, we have used $UZU^{\dagger}=Y$, $UXU^{\dagger}=X$ and that $Y_i$, being the product of $X$ and $Z$ up to a phase, is represented as $[1_i\mid1_{i+n}]$ in the symplectic representation~\cite{Aaronson2004}. Therefore, the transversal application of $U$ copies block matrices from the right $Z$-part to the left $X$-part of the symplectic stabilizer representation. By row addition (stabilizer multiplication), this state coincides with
\begin{equation}
    \begin{bmatrix}
    a_x\\
    b_x\\
    a_z\oplus a_x \oplus r\\
    b_z\oplus b_x \oplus r
    \end{bmatrix},
    \begin{bmatrix}[cc|cc]
    H_x & 0 & 0 & 0\\
    0 & H_x & 0 & 0\\
    0 & 0 & H_x & 0\\
    0 & 0 & 0 & H_x
    \end{bmatrix},
\end{equation}
where $r$ represents the row sum modulo $4$ for the matrix $H_x$ (see \textit{rowsum} in Ref.~\cite{Aaronson2004}). Analogously to the previous derivation of Eq.~\eqref{eq_binary_add}, one finds that transversal CNOT gates result in the state 
\begin{equation}
    \ket{C_{a_x\oplus b_x, a_z \oplus a_x \oplus r}}\otimes \ket{C_{b_x, b_z\oplus b_x \oplus a_z \oplus a_x}}.
\end{equation}
Afterwards, the second state is measured in the $Z$-basis. We post-select on the condition $b_z\oplus b_x \oplus a_z \oplus a_x=0$ which removes bit-flip and phase-flip errors on both states, which can be used to achieve faster convergence in cascaded purification of Bell states~\cite{Dur1999}. However, the new scheme is not generally superior. The previously discussed bit-flip (phase-flip) purification is sensitive to Pauli $X, Y$ ($Y,Z$) errors while being insensitive to Pauli $Z$ ($X$) errors. The scheme discussed here is sensitive to Pauli $X,Z$ errors while being insensitive to Pauli $Y$ errors. It thus depends on the noise type which scheme performs best.

After purification, the coefficients of the new mixed state are
\begin{align}
    \alpha_{a_x',a_z'}=&\frac{1}{\sigma}\sum_{\substack{a_x,a_z,b_x,b_z \\ a_x\oplus b_x=a_x'\\a_x\oplus a_z \oplus r=a_z'\\b_z\oplus b_x\oplus a_z \oplus a_x= 0}}\alpha_{a_x,a_z}\cdot\alpha_{b_x,b_z}\\
    =&\frac{1}{\sigma}\sum_{a_x}\alpha_{a_x,a_x\oplus a_z' \oplus r}\cdot\alpha_{a_x\oplus a_x',a_x\oplus a_x'\oplus a_z'\oplus r},
\end{align}
where the normalization factor $1/\sigma$ is the inverse of the probability of success, $\sigma$.

As before the purification scheme can be implemented using type-I fusions. The only difference to the implementation of the previous bit- and phase-flip purification schemes is that the transversal gates $U=e^{i\frac{\pi}{4}X}$ are applied before the type-I fusions. $H_x=H_z$ is necessary and sufficient for the above purification scheme. The condition implies $H_xH_x^T=0$ because all stabilizers must commute. Therefore, the matrix $H_x=H_z$ corresponds to a self-dual binary code~\cite{Macwilliams1977}. An example is the [8,4,4] extended Hamming code. The corresponding CSS state, with $H_x=H_z$ representing both $X$- and $Z$-type stabilizers, is locally equivalent to the eight qubit cube graph state~\footnote{Many more examples can be constructed by choosing the standard form $H_x=H_z=[1\mid A]$, where $A$ is an orthogonal matrix, i.e. $AA^T=1$~\cite{Macwilliams1977}.}. Indeed, this graph state represents a resource in fusion-based quantum computing~\cite{Bell2022} and thus the example shows that the outlined scheme can be applied to some two-colorable graph states of practical interest.

The condition $H_x=H_z$ is often not fulfilled, in which case the above scheme does not apply. However, it has been shown in Ref.~\cite{Hostens2006} that applying Clifford gates other than CNOT gates can be generally exploited for the purification of CSS states. Similar generalizations may also be applied when type-I fusions are used.

\section{Generalization to all stabilizer states}
\label{sec_all_graph}
Using deterministic CNOT gates, it has been shown that also graph states that are not two-colorable can be purified~\cite{Kruszynska2006}. In this section, we explain this protocol and then describe how it can be implemented with type-I fusions and provide the corresponding success probabilities.

\subsection{The general purification protocol}
\begin{figure}
\includegraphics[width=1.0\linewidth]{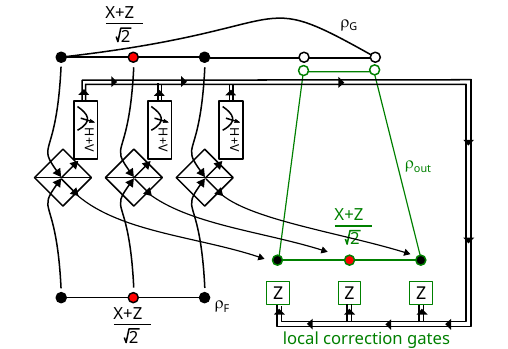}
\caption{\label{fig_all_graph}Setup for the purification of a five-ring graph state (chromatic number $3$) with density matrix $\rho_G$. For the first purification round, a three-qubit ancillary graph state with density matrix $\rho_F$ is sent into a transversal type-I fusion setup together with the first three qubits of the five-ring graph state. On the second qubits, Hadamard gates ($\frac{X+Z}{\sqrt{2}}$) are applied prior to type-I fusion. The shown purification reduces phase-flip errors on the second qubit and bit-flip errors on its neighbors. Local correction $Z$-gates are applied when the detection of the type-I fusion projects on the $\ket{-}$ basis.}
\end{figure}

The key idea for purifying arbitrary graph states is purifying only part of the state using an ancillary state which is a two-colorable subgraph of the target state~\cite{Kruszynska2006, Goyal2006}. Purification is repeated with different ancilla states until purification has improved the error rate on each qubit of the target graph state. A corresponding purification setup for partial purification of a five-ring graph state is shown in Fig.~\ref{fig_all_graph}.

The subgraph forming the ancilla state can be defined by choosing an independent set of nodes on the target graph, where independent means that the nodes are mutually not adjacent. Assume a graph $G$ with an independent set of nodes $A$. We denote the $1$-neighborhood of $A$, i.e., all nodes that are not in $A$ but adjacent to at least one node in $A$, as $\bar{A}$. The set of all other nodes is denoted as $B$. In block-matrix representation, the associated graph basis state~\cite{Looi2008} $\ket{
G_{g_A,g_{\bar{A}},g_B}}$ has the symplectic representation
\begin{equation}
    \begin{bmatrix}
    g_A\\
    g_{\bar{A}}\\
    g_B
    \end{bmatrix},
    \begin{bmatrix}[ccc|ccc]
    1_A & 0  & 0 & 0 & R & 0\\
    0 & 1_{\bar{A}} & 0 & R^T & Q & S\\
    0 & 0 & 1_B & 0 & S^T & T
    \end{bmatrix},
\end{equation}
where $1_A, 1_{\bar{A}}, 1_B$ represent identity matrices and $Q, R, S, T$ are blocks of the adjacency matrix of $G$.

The corresponding ancillary graph (basis) state $\ket{
F_{g_A,g_{\bar{A}}}}$ has the symplectic representation
\begin{equation}
    \begin{bmatrix}
    f_A\\
    f_{\bar{A}}\\
    \end{bmatrix},
    \begin{bmatrix}[cc|cc]
    1_A & 0 & 0 & R\\
    0 & 1_{\bar{A}} & R^T & 0\\
    \end{bmatrix}.
\end{equation}
Since this ancillary state is two-colorable, it can be purified independently with the protocol of Section~\ref{sec_css}. It corresponds to the subgraph of $G$ with the nodes of $A$ and the nodes of its $1$-neighborhood $\bar{A}$, with only edges between $A$ and $\bar{A}$.

Applying transversal $H$-gates to the qubits in set $A$ of both states transforms $\ket{G_{g_A,g_{\bar{A}},g_B}}\otimes\ket{F_{g_A,g_{\bar{A}}}}$ into
\begin{equation}
    \label{eq_general_init}
    \begin{bmatrix}
    g_A\\
    g_{\bar{A}}\\
    g_B\\
    f_A\\
    f_{\bar{A}}\\
    \end{bmatrix},
    \begin{bmatrix}[ccccc|ccccc]
    0 & 0 & 0 & 0 & 0 & 1_A & R & 0 & 0 & 0\\
    R^T & 1_{\bar{A}} & 0 & 0 & 0 & 0 & Q & S & 0 & 0\\
    0 & 0 & 1_B & 0 & 0 & 0 & S^T & T & 0 & 0\\
    0 & 0 & 0 & {\color{blue}0} & {\color{blue}0} & 0 & 0 & 0 & {\color{blue}1_A} & {\color{blue}R}\\
    0 & 0 & 0 & {\color{blue}R^T} & {\color{blue}1_{\bar{A}}} & 0 & 0 & 0 & {\color{blue}0} & {\color{blue}0}\\
    \end{bmatrix},
\end{equation}
where the columns/rows representing $H^{\otimes A}\ket{F_{g_A,g_{\bar{A}}}}$ are highlighted in blue.

We apply transversal CNOT gates with the qubit in $A,\bar{A}$ of $H^{\otimes A}\ket{G_{g_A,g_{\bar{A}},g_B}}$ as the control and the corresponding qubits in the ancilla $H^{\otimes A}\ket{F_{g_A,g_{\bar{A}}}}$ as the targets, followed by measuring the ancilla qubits in the $Z$-basis. The action of the transversal CNOT gates leads to
\begin{equation}
    \begin{bmatrix}
    g_A\\
    g_{\bar{A}}\\
    g_B\\
    f_A\\
    f_{\bar{A}}\\
    \end{bmatrix},
    \begin{bmatrix}[ccccc|ccccc]
    0 & 0 & 0 & 0 & 0 & 1_A & R & 0 & 0 & 0\\
    R^T & 1_{\bar{A}} & 0 & R^T & 1_{\bar{A}} & 0 & Q & S & 0 & 0\\
    0 & 0 & 1_B & 0 & 0 & 0 & S^T & T & 0 & 0\\
    0 & 0 & 0 & 0 & 0 & 1_A & R & 0 & 1_A & R\\
    0 & 0 & 0 & R^T & 1_{\bar{A}} & 0 & 0 & 0 & 0 & 0
    \end{bmatrix},
\end{equation}
which is identical by row addition (stabilizer multiplication) to
\begin{equation}
    \label{eq_general_final}
    \begin{bmatrix}
    g_A\\
    g_{\bar{A}}\oplus f_{\bar{A}}\\
    g_B\\
    f_A \oplus g_A\\
    f_{\bar{A}}\\
    \end{bmatrix},
    \begin{bmatrix}[ccccc|ccccc]
    0 & 0 & 0 & 0 & 0 & 1_A & R & 0 & 0 & 0\\
    R^T & 1_{\bar{A}} & 0 & 0 & 0 & 0 & Q & S & 0 & 0\\
    0 & 0 & 1_B & 0 & 0 & 0 & S^T & T & 0 & 0\\
    0 & 0 & 0 & 0 & 0 & 0 & 0 & 0 & 1_A & R\\
    0 & 0 & 0 & R^T & 1_{\bar{A}} & 0 & 0 & 0 & 0 & 0
    \end{bmatrix}.
\end{equation}
The measurement of the qubits of the ancilla state in the $Z$-basis leads to the state 
\begin{equation}
    \begin{bmatrix}
    g_A\\
    g_{\bar{A}}\oplus f_{\bar{A}}\\
    g_B
    \end{bmatrix},
    \begin{bmatrix}[ccc|ccc]
    0 & 0 & 0 & 1_A & R & 0\\
    R^T & 1_{\bar{A}} & 0 & 0 & Q & S\\
    0 & 0 & 1_B & 0 & S^T & T
    \end{bmatrix},
\end{equation}
which is $H^{\otimes A}\ket{G_{g_A,g_{\bar{A}}\oplus f_{\bar{A}},g_B}}$. At the same time, all stabilizers from the fourth row in Eq.~\eqref{eq_general_final} can be reconstructed from the measurement outcomes, allowing post-selecting on $f_A\oplus g_A = 0$. If the initial states $\ket{G_{g_A, g_{\bar{A}},g_B}}$ and $\ket{F_{g_A, g_{\bar{A}}}}$ have imperfections, this post-selection enables first-order filtering of phase-flip errors of qubits in the independent set $A$ and bit-flip errors in $\bar{A}$.

Assume the initial target state has the form
\begin{equation}
\rho_G=\sum_{g_A,g_{\bar{A}},g_B}\alpha_{g_A,g_{\bar{A}},g_B}\ket{G_{g_A,g_{\bar{A}},g_B}}\bra{G_{g_A,g_{\bar{A}},g_B}}
\end{equation}
and the mixed ancillary state has the form
\begin{equation}
\rho_F=\sum_{f_A,f_{\bar{A}}}\Tilde{\alpha}_{f_A,f_{\bar{A}}}\ket{F_{f_A,f_{\bar{A}}}}\bra{F_{f_A,f_{\bar{A}}}}.
\end{equation}
Applying $H^{\otimes A}$ after successful purification, the new state has the coefficients~\cite{Kruszynska2006}
\begin{equation}
    \alpha_{g_A',g_{\bar{A}}',g_B'} = \frac{1}{\sigma}\sum_{g_{\bar{A}},f_{\bar{A}}: (g_{\bar{A}}\oplus f_{\bar{A}}) = g_{\bar{A}}'} \alpha_{g_A'g_{\bar{A}}g_B'}\Tilde{\alpha}_{g_A'f_{\bar{A}}},
\end{equation}
where $\sigma$ normalizes the new probability distribution and corresponds to the probability that the purification is successful.

\subsection{Realization with type-I fusions}
As we have discussed before, successful type-I fusion has, up to a heralded correction gate, the effect of a CNOT gate followed by a measurement of the target in the $Z$-basis. Therefore, we can implement the above purification protocol without deterministic CNOT gates and by employing type-I fusions. Type-I fusions are applied between qubits $A, \bar{A}$ from the target state $H^{\otimes A}\ket{G_{g_A,g_{\bar{A}},g_B}}$ and the corresponding qubits from the ancilla state $H^{\otimes A}\ket{F_{g_A,g_{\bar{A}}}}$. The probability that all type-I fusions succeed is given by the term
\begin{equation}
    \label{eq_success_all}
    p_s=\sigma/2^{\mid\bar{A}\mid},
\end{equation}
where $\abs{\bar{A}}$ is the cardinality of the set $\bar{A}$. The factor $\sigma$ represents the probability that the purification does not fail due to error detection and is one for a perfect input state. The factor $1/2^{\mid \bar{A} \mid}$ is because the $\abs{\bar{A}}+\abs{A}$ type-I fusions are deterministic $\abs{A}$ times and probabilistic $\abs{\bar{A}}$ times, which is due to an argument very similar to Section~\ref{sec_css}: in Eq.~\eqref{eq_general_init}, adding the fourth row on the first, shows that $\pm Z_{A_{g,i}}Z_{A_{f,i}}$ is a stabilizer for $\abs{A}$ different rows $i$, making the type-I fusions deterministic $\abs{A}$ times. Considering the second and fifth row, one can see that there is a minimum number of $\abs{\bar{A}}$ stabilizer generators with support on $X_{\bar{A}_{g,i}}\mathds{1}_{\bar{A}_{f,i}}$ or $\mathds{1}_{\bar{A}_{g,i}}X_{\bar{A}_{f,i}}$, making the type-I fusions probabilistic $\abs{\bar{A}}$ times.

\subsection{Multiple purification rounds}
The above purification protocol filters bit- and phase-flip errors in a subset of the qubits of the target state. Therefore, more than one purification rounds are required for a full purification.

The question of how many rounds of the described purification protocol are required is linked to graph-colorability. Given a $k$-coloring of the graph $G$, the set $A$ can be chosen as one of the colors as every color makes an independent set. Repeating the protocol in $k$ rounds, with $A$ corresponding to a different color in each round, will ensure that for every qubit, phase-flip errors are filtered once~\cite{Kruszynska2006}. This is because the $k$-coloring represents a partition of the graph and so every node is in the set $A$ once. At the same time, bit-flip errors are filtered at least once for every qubit. This is because for a connected graph with more than one node, every node is in the $1$-neighborhood of at least one other color. If $k$ corresponds to the chromatic number of $G$, the graph cannot be partitioned into fewer independent sets. The chromatic number therefore determines the minimum number of cascaded purification rounds required to obtain first-order tolerance to a single-qubit error on any of the qubits.

Furthermore, we note that the set $A$ must be independent, but does not need to correspond to a color in a graph coloring that uses the minimum number of colors. A different choice of an independent set can make sense if only a single purification round can be performed, and thus one intends to maximize $A$. Indeed, there exist graphs for which the largest independent set is larger than the largest set of nodes with identical color in any minimum-color graph coloring. An example is the graph with edges $(0,1), (1,2), (2,3), (1,4), (2,5)$ that is two-colorable but its largest independent set $\{0,3,4,5\}$ is part of three-colorings but not part of any two-coloring. In another extreme case (like in Fig.~\ref{fig_all_graph}), the set $A$ is just one qubit and the ancilla state is thus a star graph state (locally equivalent to a GHZ state). If the valency of the graph state, $\ket{G}$, to be purified is low, the ancilla state is small and thus can be independently purified with low overhead such that its error rates are far below those of $\ket{G}$. In this case, the ancilla would introduce almost no additional errors.

Finally, when type-I fusions are used, the dependence of the success probability on $\abs{\bar{A}}$ in Eq.~\eqref{eq_success_all} suggests some optimization problems. First, assume that a graph with chromatic number $k$ is purified in $k$ rounds. In each of the required $k$ purification rounds $A$ is chosen as one color of a $k$-coloring. To minimize the multiplexing overhead, one intends to maximize the probability of success in each round. One strategy to achieve this would be to find the $k$-coloring for which the sum of the cardinality of all $1$-neighborhoods $\bar{A}$ is minimum. Second, the local equivalence class of a graph state (also called graph orbit~\cite{Adcock2020, Cabello2011}) may contain several graph states with the same chromatic number. The success probability may be highest when purifying one particular state, likely one with a low average vertex degree.

\section{Generating and purifying time-bin-encoded states}
\label{sec_time_bin}
\subsection{Setup and Protocol}
In Fig.~\ref{fig_setups}, we have shown purification setups for polarization encoding and path encoding. Instead of polarization-encoded photonic graph states~\cite{Schwartz2016, Cogan2023, Huet2026}, time-bin-encoded graph states can be generated with quantum emitters~\cite{Tiurev2021, Tiurev2021b} and may have several advantages: (1) Time-bin encoding is more robust in optical fibers. (2) The corresponding graph state generation scheme operates at high magnetic fields~\cite{Meng2023b}. For quantum dots serving as quantum emitters, the noise on the hole spin is reduced~\cite{Huthmacher2018} at high magnetic fields, and nuclear spin-cooling via the electron spin is possible~\cite{Gangloff2019}.

\begin{figure*}[!t]
\includegraphics[width=1.0\textwidth]{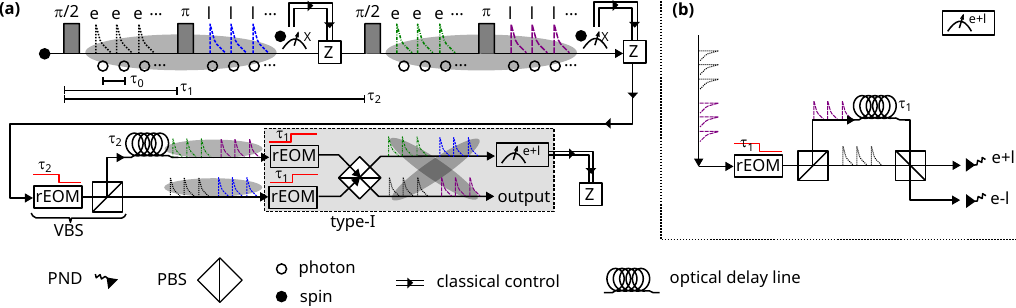}
\caption{\label{fig_time_bin}\textbf{(a)} Setup for generation and bit-flip purification of time-bin-encoded GHZ states (e.g. $\ket{\Phi_0^+}$). The letters $e, l$ represent the early and late time bins, which are the computational basis states. Compared to polarization or path encoding, the transversal type-I fusions can be implemented using fewer spatial or polarization modes by putting several qubits into the same mode. Moving to time-bin encoding, the polarizing beam splitters from Fig.~\ref{fig_setups}(a) (polarization encoding) are replaced by variable beam splitters (VBS), implemented by a resonant electro-optic modulator (rEOM) and a polarizing beam splitter (PBS). The VBS transmits the early time bin and reflects the late one. \textbf{(b)} Setup for detecting a time-bin-encoded qubit in the superposition basis $(\ket{e+l}, \ket{e-l})$.}
\end{figure*}

Fig.~\ref{fig_time_bin}(a) shows a purification setup (first round, bit-flip correction) for time-bin-encoded three-qubit GHZ states, generated by a single quantum emitter with a spin~\cite{Lindner2009, Tiurev2021, Tiurev2021b}. First, the spin is brought to the state $\frac{1}{\sqrt{2}}\left(\ket{\downarrow}+\ket{\uparrow}\right)$ with a $\pi/2$ rotation pulse. Following Ref.~\cite{Tiurev2021}, we assume a level scheme where the system can be selectively excited with an optical pulse, given that the spin is in the state $\ket{\downarrow}$. Spontaneous decay of the excited state results in single-photon emission. Applying $N$ optical excitation pulses, followed by a $\pi$-pulse ($X$ gate) on the spin, followed by additional $N$ optical excitation pulses thus implements the operation $\frac{1}{\sqrt{2}}\left(\ket{\downarrow}+\ket{\uparrow}\right) \rightarrow \frac{1}{\sqrt{2}}\left(\ket{\downarrow}\ket{e}^{\otimes N}+\ket{\uparrow}\ket{l}^{\otimes N}\right)$. Here, $e, l$ represent the early and late time bins of the $N$ time-bin-encoded photonic qubits. Measuring the spin in the Pauli $X$-basis, i.e. the ($\ket{\downarrow+\uparrow},\ket{\downarrow-\uparrow}$)-basis, results in the state $\frac{1}{\sqrt{2}}\left(\ket{e}^{\otimes N}+(-1)^m\ket{l}^{\otimes N}\right)$, where $m=0,1$ depends on the outcome of the spin measurement. In the case of $m=1$, a $Z$ gate is applied before the subsequent purification (feedforward), which can be implemented by an EOM making a $\pi$-phase-shift on an odd number of late time bins. In the generated state, the time bins of a single qubit are fragmented such that the separation $\tau_1$ between the early and late time bins of the qubit is larger than the separation $\tau_0$ between time bins of subsequent qubits (see Fig.~\ref{fig_time_bin}(a)). This could be avoided by applying additional spin control pulses~\cite{Tiurev2021, Tiurev2021b}, but has the disadvantage of introducing more spin errors~\cite{Meng2023b}.

The above scheme is performed twice to generate two copies of the same state for subsequent purification. The two time-bin-encoded states are  separated by a time delay $\tau_2$. This delay is compensated by a variable beam splitter (VBS) which sends the earlier state to a corresponding delay line. The variable beam splitter can for instance be implemented by a resonant electro-optical modulator (rEOM) modulating the polarization of the photons, followed by a polarizing beam splitter~\cite{Chan2025}. Subsequently, transversal type-I fusion is achieved by using one rEOM per state and a common PBS. The rEOMs set the polarization such that the early-time-bin photons are transmitted while the late-time-bin photons are reflected by the PBS. Finally, the photons in one optical mode are measured in the $(\ket{e+l}, \ket{e-l})$-basis, as illustrated in Fig.~\ref{fig_time_bin}(b). Depending on the outcome of the measurements, a correcting Pauli $Z$-gate may need to be applied in feedforward to obtain the desired state. The correction gate is required if the state is sent to further cascaded purifications. It can be omitted by Pauli frame tracking if the state is used directly for fusion-based quantum computing.

\subsection{Rate of purified GHZ source}
Here, we estimate at which rate a single quantum dot could generate purified GHZ states. First, we consider the duration $\tau_1$ between the $\pi/2$ and the $\pi$ pulses in Fig.~\ref{fig_time_bin}(a). For quantum dots, the Hahn echo visibility showed a peak at $\tau_1=29\ \text{ns}$ in Ref.~\cite{Meng2023b} and we therefore assume this time delay. Within this time window, there must be some time separation $\tau_0$ between all optical pulses to be able to select individual time-bins with fast switches. The quantum dot trion state has a sub nano-second radiative lifetime that is typically Purcell enhanced in waveguides~\cite{Dalgarno2008, Lodahl2015}. Therefore, a time separation of $\tau_0\sim 2\ \text{ns}$ should be feasible, in which case one could generate a GHZ state with up to $\sim \lfloor29\ \text{ns} / 2\ \text{ns} \rfloor= 14$ photons. After generating a spin-photon GHZ state, the spin is measured in the $X$-basis. For the time of this measurement, we assume $10\ \text{ns}$. This is a conservative estimate and is about three times longer than what was achieved in Ref.~\cite{Antoniadis2023}. Finally, we note that the nuclear spin diffusion time is on the order of tens of milliseconds, and nuclear spin cooling can be achieved in tens of microseconds~\cite{Gangloff2019}. Therefore, nuclear spin cooling will only be applied during a fraction of the overall time, while most of the time is used for generating GHZ states.

For the purification protocol in Fig.~\ref{fig_time_bin}(b), two GHZ states are subsequently generated by the same quantum emitter, increasing the duration by a factor of two. Furthermore, purification only succeeds with a finite probability (up to $p_s^{(b)}=1/2$ for bit-flip and $p_s^{(p)}=1/2^{n-1}$ for phase-flip purification) which further reduces the rate. In summary, we obtain a rate between $p_s^{(b/p)}/2/\left(2\times29\ \text{ns} + 10\ \text{ns}\right)$ and $p_s^{(b/p)}/2/\left(29\ \text{ns} + 10\ \text{ns}\right)$. For generating a purified 3-qubit GHZ-state with a single quantum dot, this would correspond to a rate between $3.7- 6.4\ \text{MHz}$ for bit-flip purification and between $1.8-3.2\ \text{MHz}$ for phase-flip purification.

\section{Numerical Results}
\label{sec_numerics}
In this section, we numerically determine for a few states and various noise parameters the optimum purification scheme and the achievable fidelities. Such an analysis can guide the experimental realization of purification schemes where the noise rates are fixed system parameters, and one intends to choose the best possible scheme. We will first consider depolarization noise, as well as phenomenological bit- and phase-flip errors, and then physical noise sources relevant when generating time-bin-encoded caterpillar tree graph states~\cite{Tiurev2021, Tiurev2021b}.

\subsection{CSS states with depolarization and phenomenological noise}
\begin{figure}
\includegraphics[width=1.0\linewidth]{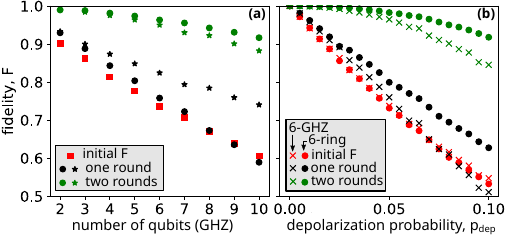}
\caption{\label{fig_fidelity}\textbf{(a)} Fidelities of GHZ states of various sizes after zero (red), one (black), and two (green) purification rounds, assuming depolarization noise with $p_{dep}=0.05$. For one round, we simulate bit- and phase-flip purification. For two rounds, we consider two cases: one where the phase-flip purification is performed before bit-flip purification (circles), and one where the bit-flip purification is performed before phase-flip purification (stars). \textbf{(b)} Initial fidelities, after one round of phase-flip purification, and after two rounds of purification (phase-flip, followed by bit-flip purification) as a function of $p_{dep}$. The simulation is performed for a six-qubit GHZ state (crosses) and for a CSS state corresponding to a six-qubit ring graph state (circles).}
\end{figure}

First, we consider the purification of GHZ states of different sizes. In this case, Eqs.~\eqref{eq_purify1} and~\eqref{eq_purify_2} determine the state after one purification round. First, we consider depolarization noise parameterized by a probability $p_{dep}=0.05$, with $p_{dep}/3$ being the probability of Pauli $X,Y,Z$ errors, respectively. Fig.~\ref{fig_fidelity}(a) shows the corresponding fidelities before purification, after one round of purification, and after two rounds. One can see that two purification rounds significantly improve the fidelity (for instance, from $F=0.81$ to $F=0.98$ for a four-qubit GHZ state). We find that the improvement of the fidelities is much stronger when using two purification rounds rather than just a single one. The reason why this makes such a big difference is that a single purification round does not correct all error types to first order (see Fig.~\ref{fig_illustration}), whereas two rounds do.

We find that the highest fidelities are achieved when phase-flip purification is performed before bit-flip purification. When the two purification rounds are performed in opposite order, the fidelity is lower, reaching only $F=0.97$ for the four-qubit GHZ state. This difference is not obvious, and which order is best can be different for different noise parameters.

\begin{figure}[!t]
\includegraphics[width=1.0\linewidth]{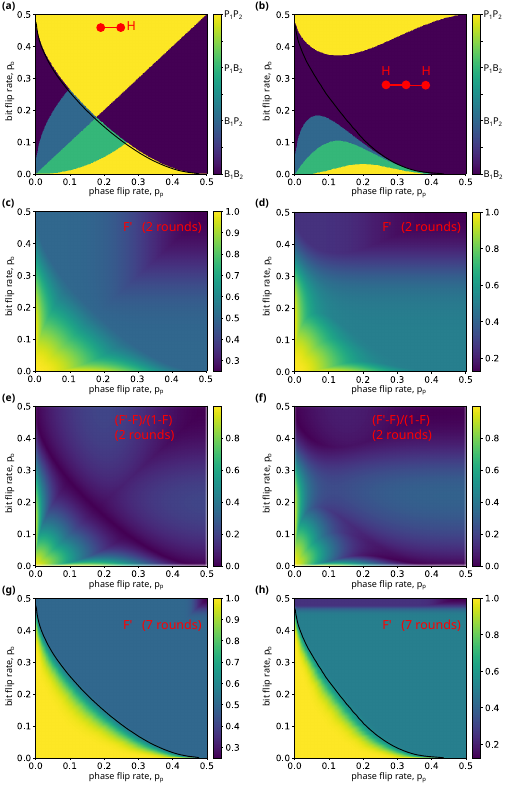}
\caption{\label{fig_results_GHZ} \textbf{(a, b)} Best GHZ state purification protocols for phenomenological bit- and phase-flip noise, allowing for two rounds of bit-flip (B) and phase-flip (P) purifications. The results are shown in (a) for the two-photon Bell state $\frac{1}{\sqrt{2}}\left(\ket{00}+\ket{11}\right)$ and in (b) for the three-photon GHZ state $\frac{1}{\sqrt{2}}\left(\ket{000}+\ket{111}\right)$. \textbf{(c,d)} Corresponding fidelities $F'$ after the two purification rounds. \textbf{(e, f)} Relative amount of removed infidelity, i.e. $(F'-F)/(1-F)$ corresponding to the two purification rounds above. \textbf{(g, h)} Highest fidelities that can be achieved with seven cascaded purification rounds.}
\end{figure}

An interesting question is therefore in which parameter regime which combination of phase-flip and bit-flip corrections leads to the best fidelity $F'$ of the final state. To investigate this, we apply phenomenological and independent bit-flip and phase-flip error rates of $p_b, p_p$ to GHZ states, and assume that two cascaded purification rounds can be performed (either bit- or phase-flip purification).  Under this constraint, Fig.~\ref{fig_results_GHZ}(a, b) shows the optimum purification protocols for two- and three-qubit GHZ states. For the two qubit GHZ (Bell) state, this simulation only has an illustrative purpose since the improved purification scheme from Section~\ref{sec_dejmps} is not considered. In Fig.~\ref{fig_results_GHZ}(c, d) we plot the corresponding fidelities achieved after purification, and the relative amount of removed infidelity is shown in Fig.~\ref{fig_results_GHZ}(e, f). (If the initial state had a fidelity of $F=0.9$ and the purified state had a fidelity of $F'=0.95$, the relative amount of removed infidelity would be $(F'-F)/(1-F)=0.5$). Which purification protocol is best non-trivially depends on the noise parameters (see Figs.~\ref{fig_results_GHZ}(a,b)), but you can get a certain sense of it: for instance, when there are only phase (bit) flip errors, the simulations show that it is advantageous to do cascaded phase(bit)-flip purification rounds without any correction of the other noise type. Furthermore, for the three-qubit GHZ state, it is advantageous to correct for only bit-flip errors in a large part of the parameter regime. This effect becomes stronger for states of larger size, illustrating that for GHZ states bit-flip errors can be removed more easily than phase flip errors due to their repetition-code structure.

At low bit- and phase-flip error rates, the optimum purification protocol strongly improves fidelities. We also find numerically that, when selecting the optimum purification scheme, purification always improves the fidelities, and adding more purification rounds is always advantageous. However, there are certain parameter regimes in which purification is challenging. First, the transition between different purification protocols in Fig.~\ref{fig_results_GHZ}(a,b) can be seen as a faint reduction in the achieved fidelity (Fig.~\ref{fig_results_GHZ}(c,d)) and the fraction of removed infidelity (Fig.~\ref{fig_results_GHZ}(e,f)). Second, it turns out that there is a threshold that marks a limit above which the fidelity of the state cannot be improved to unity. This is akin to bound entanglement~\cite{Horodecki1998} that cannot be distilled~\footnote{Bound entanglement is typically considered in the context of general purification protocols, while we investigate a particular class of purification protocols.}. The effect can be seen in Fig.~\ref{fig_results_GHZ}(g, h), where we allow up to seven cascaded purification rounds and plot the fidelity maximized over all possible sequences of bit- and phase-flip purification rounds. At low bit- and phase-flip rates, the achieved fidelity approaches unity, but there is a transition above which the fidelity converges to a plateau value of only 0.5, marking a threshold for purification. As the number of purification rounds increases, bit-flip errors are progressively removed; however, when the noise is too large, the protocol drives the state toward an equal mixture of $\ket{\Phi_0^{+}}$ and $\ket{\Phi_0^{-}}$. This explains the fidelity reaching the limit of 0.5 and again illustrates that bit-flip errors are more effectively suppressed than phase-flip errors for GHZ states. In Fig.~\ref{fig_results_GHZ}(a,b,g,h), the faint black lines show where, for the seven optimized cascaded purification rounds, the drop in fidelity from unity to the plateau value reached $90\%$ (a fidelity drop to 0.55 for a plateau of 0.5). The line thus represents a lower bound for the purification threshold. For states such as the Bell state in Fig.~\ref{fig_results_GHZ}(a) or the CSS states locally equivalent to the cube graph state, we have performed additional simulations allowing the scheme from Section~\ref{sec_dejmps} as a third purification scheme in every round. In the correctable regime, this makes the convergence towards unity fidelity faster but seems to have no effect on the threshold itself.

\begin{figure}
\includegraphics[width=1.0\linewidth]{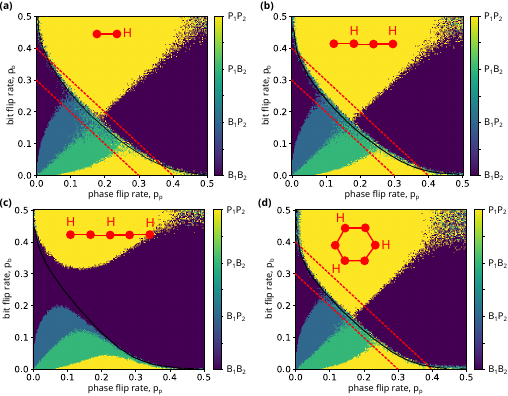}
\caption{\label{fig_results_CSS} Best CSS state purification protocols for phenomenological bit- and phase-flip noise and two rounds of bit-flip (B) and phase-flip (P) purification. The simulated CSS states are locally equivalent by Hadamard (H) gates to the following graph states: \textbf{(a)} a length-two linear chain, \textbf{(b)} a length-four linear chain, \textbf{(c)} a length-five linear chain, and \textbf{(d)} a six-ring. In all plots, the faint black lines indicate a lower bound for the purification threshold. In (a,b,d), the dotted red lines are a guide to the eye to see the slight difference between the simulations.}
\end{figure}

Next, we consider the purification of more general CSS states, i.e. states locally equivalent to two-colorable graph states. For two exemplary CSS states and depolarization noise, Fig.~\ref{fig_fidelity}(b) shows the fidelities before purification, after one round of phase-flip purification, and after two purification rounds (phase-flip followed by bit-flip purification). One state is the CSS state locally equivalent to the six-qubit ring graph state, and the other state is the six-qubit GHZ state. In this example, we find that the GHZ state can have lower fidelity after one round of purification, indicating the high noise-sensitivity of GHZ states.

Fig.~\ref{fig_results_CSS} shows the optimum two-round purification protocols for a few exemplary CSS states: linear chains of lengths two, four, and five, as well as a ring graph state with six qubits. First, we note that the results from Fig.~\ref{fig_results_CSS}(a) and Fig.~\ref{fig_results_GHZ}(a) coincide, which is expected since a length-two linear chain graph state with one $H$-gate is a two-qubit GHZ (Bell) state. The noise in the corresponding CSS simulation is due to the fact that the initial density matrix is estimated by Monte Carlo sampling with $3\times 10^4$ samples~\footnote{This number of samples is also used in subsequent simulations.}, randomly applying bit- and phase-flips. For the GHZ simulation, we instead determine the probabilities of the initial density matrix analytically. Furthermore, we observe that the simulations in Fig.~\ref{fig_results_CSS}(a,b,d) are mirror symmetric about the diagonal. This behavior is expected, as the corresponding CSS states have identical Pauli $X$- and Pauli $Z$-type stabilizers. The simulation results for the three different states are very similar, but a slight difference can be seen by comparing the area enclosed by the dotted red lines in Fig.~\ref{fig_results_CSS}. For the length-five linear chain in Fig.~\ref{fig_results_CSS}(c), the symmetry between phase- and bit-flips is not present, which is the case for all chains with an uneven number of qubits.

\subsection{CSS states with physical noise}
The above analysis is based on a simplified error model. In physical graph state generation, several more subtle noise sources can affect fidelities, and a catalog of error sources for polarization and time-bin encoding has been presented in Refs.~\cite{Chan2025, Sheldon2025, Prasad2025}. Here, we consider two error sources that arise when generating time-bin-encoded graph states with quantum emitters~\cite{Tiurev2021b, Tiurev2021} (see Section~\ref{sec_time_bin}, Fig.~\ref{fig_time_bin}). First, we consider a characteristic branching error arising from a finite cyclicity of the used level scheme (see Fig. 8a in Ref.~\cite{Chan2025}). This error is parameterized by the branching probability $p_{br}$, specifying the probability that the optically excited state in the early time bin decays via an undesired transition to the the wrong spin state. Second, we consider a finite distinguishability $1-V$ between the emitted photons. We assume here that the distinguishability errors of all photons in both input states are not correlated, e.g. corresponding to the decoherence caused by a rapidly varying phonon reservoir. More error sources such as the finite spin coherence time $T_2$ could be investigated.

\begin{figure}[ht!]
\includegraphics[width=1.0\linewidth]{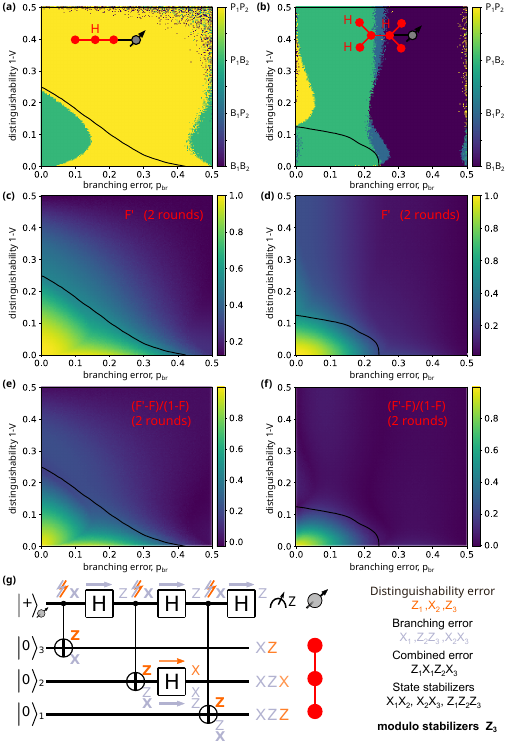}
\caption{\label{fig_results_physical_noise}\textbf{(a, b)} Best purification protocols for physical noise in time-bin-encoded graph state generation~\cite{Tiurev2021}. Photons are generated in the time order from left to right and $H$-gates are applied afterwards to some photonic qubits, turning the graph state into a CSS state. Noise arises from a combination of finite distinguishability and a finite cyclicity~\cite{Chan2024}. Two cascaded rounds of bit-flip (B) and phase-flip (P) corrections are allowed. The faint black lines indicate a lower bound of the threshold below which purification to arbitrary high fidelities is possible (based on a simulation of all possibilities for seven cascaded purification rounds). \textbf{(c, d)} Corresponding fidelities $F'$ after the two purification rounds. \textbf{(e, f)} Relative amount of removed infidelity $(F'-F)/(1-F)$. \textbf{(g)} Generation circuit of the 3-photon state from (a), together with Pauli errors (bold X/Z letters) originating from branching errors (light blue) and distinguishability (orange) in time-bin-encoded graph state generation~\cite{Tiurev2021, Chan2025}. To explain the dominant error character qualitatively, we consider the maximum errors possible from both noise sources, their error propagation, and determine the resulting error string modulo state stabilisers.}
\end{figure}

We consider CSS states that are locally equivalent to redundantly encoded linear chain graph states. More specifically, these states are locally equivalent to caterpillar tree graph states with all nodes of the central path having the same vertex degree. For sampling errors on such states, we apply the source code from Ref.~\cite{Chan2025}, with two minor modifications: first, the spin is measured in the $Z$-basis at the end of the graph-state generation sequence to detach it from the photonic state. Second, we include the $H$-gates that convert the encoded linear chain graph states into the corresponding CSS states.

In Fig.~\ref{fig_results_physical_noise}(a, b), we show the optimum purification protocol for two CSS states under the above noise model. The achieved fidelities and the fraction of removed infidelities are shown in Fig.~\ref{fig_results_physical_noise}(c, d) and Fig.~\ref{fig_results_physical_noise}(e, f), respectively. Again, the best purification protocol is highly dependent on the state and the noise parameters. In Fig.~\ref{fig_results_physical_noise}(a), two consecutive phase-flip purifications ($P_1P_2$) are often preferred because for this three-photon CSS state the physical noise maps predominantly onto $Z$-type errors (reduced modulo the stabilizer group). An example of this effect is illustrated in Fig.~\ref{fig_results_physical_noise}(g) for the three-qubit chain, where we consider the maximum set of Pauli errors from both finite indistinguishability and branching, perform error propagation, and identify equivalent errors under the stabilizers of the state~\cite{Chan2024}. However, in a substantial part of the regime where $P_1P_2$ is preferred, unity fidelity cannot be reached (the black lines in Fig.~\ref{fig_results_physical_noise}(a-e) indicate a lower-bound estimate of the boundary between the two regimes). At low physical error rates, where purification towards unity fidelity is possible with multiple concatenated purification rounds, the protocol $P_1B_2$ can become preferable: an initial phase-flip round removes the dominant $Z$-component, after which a residual $X$-component can limit the infidelity and is efficiently suppressed by a second bit-flip purification round. 

In Fig.~\ref{fig_results_physical_noise}(b), we consider the best purification protocol for a larger state. First, the plot shows that bit-flip purification is most effective in the high branching error rate regime, which is expected since the branching error flips the spin. We find with the type of analysis from Fig.~\ref{fig_results_physical_noise}(g) that the Hadamard gates convert the $Z$-heavy distinguishability contribution into an $X$-heavy one, making two-fold bit-flip purification ($B_1B_2$) more effective, even at high distinguishability contribution.  At lower branching error rates, there is a partial cancellation of $X$ errors from the two noise sources, leading to more $Z$ errors. For this reason, concatenated bit- and phase-flip purification ($P_1B_2$) is best in the practically interesting regime where the state can be purified towards unity fidelity. We remark that the photon CSS state in Fig.~\ref{fig_results_physical_noise}(b) has identical $X$- and $Z$-type stabilizers and one thus may expect a symmetry about the diagonal similar to Fig.~\ref{fig_results_CSS}(a). However, the considered graph-state generation protocol~\cite{Lindner2009, Tiurev2021} creates photons from left to right which breaks this symmetry.

\begin{figure}
\includegraphics[width=1.0\linewidth]{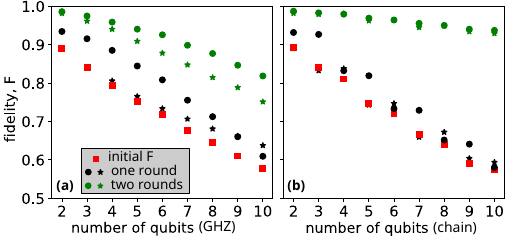}
\caption{\label{fig_fidelity_real_noise}Fidelities of GHZ states \textbf{(a)} and linear chain graph states \textbf{(b)} of various sizes after zero (red), one (black), and two (green) purification rounds. Physical noise sources corresponding to $V=0.97$ and $p_{br}=1/36$ are assumed. Like in Fig.~\ref{fig_fidelity}(a), we consider performing either phase-flip purification first (circles), or performing bit-flip purification first (stars).}
\end{figure}

In Fig.~\ref{fig_fidelity_real_noise}, we show how the fidelities of GHZ and linear chain graph states of different sizes improve upon purification. For the simulations, we have assumed an indistinguishability of $V=0.97$ as achieved in Ref.~\cite{Loredo2025} and we have assumed $p_{br}=1/36$ which is one over the cyclicity achieved in Ref.~\cite{Meng2023b}. Linear chains are clearly easier to purify than GHZ states under this noise model, but the fidelity is substantially improved by the two cascaded purification rounds in both cases. This is an encouraging result, as it shows that the purification of quantum-emitter-generated states can strongly improve fidelities.

\subsection{Purification of general graph states suffering phenomenological noise}
\begin{figure}
\includegraphics[width=1.0\linewidth]{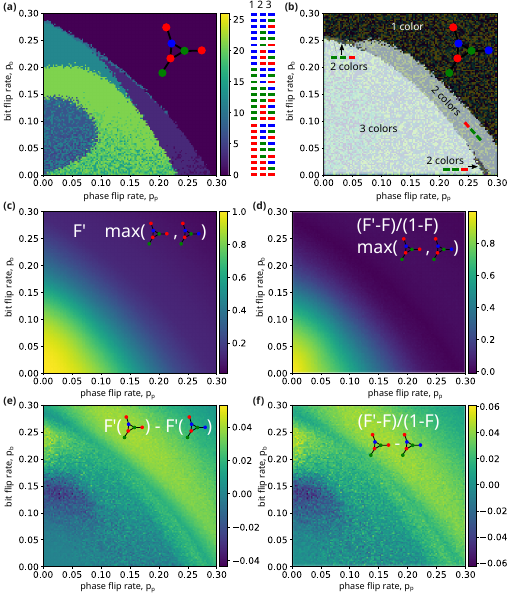}
\caption{\label{fig_results_all} Best graph state purification scheme for a graph state corresponding to a graph that is not two-colorable. The graph has two non-isomorphic 3-colorings which are independently simulated in subplots \textbf{(a)} and \textbf{(b)}. In three subsequent purification rounds, the choice of the independent set $A$ follows the graph colorings in the inset. Following a triple number system, the order in which the colors are chosen for the purification is indicated by the color bar. The colored graph in (b) is isomorphic under color swapping and therefore several purification protocols are equivalent. Unique purification protocols have been assigned the same gray scale. \textbf{(c)} Best fidelity $F'$ that can be obtained after all three purification rounds, maximizing over the two colorings and the respective order of the colors. \textbf{(d)} Fraction of removed infidelity corresponding to the plot (c). \textbf{(e)} Difference of best fidelities for the two different color choices. \textbf{(f)} Corresponding difference of the fractions of removed infidelities.}
\end{figure}

Finally, we consider the purification of graph states which are not two-colorable. We again consider phenomenological bit- and phase-flip noise with identical error rates for the initial graph state as well as the employed ancillary graph state. We assume that the employed two-colorable ancillary graph states are purified in advance in two purification rounds. Furthermore, we allow exactly $k$ purification rounds for a graph of chromatic number $k$ and choose the independent sets $A$ according to one of the corresponding graph colorings~\footnote{We find that up to eight qubits, there is no graph equivalence class where all graphs have a chromatic number larger than three. Therefore, we have $k=3$ for all simulations.}. There may be various $k$ colorings and we have performed independent simulations for colorings that are non-isomorphic. For an exemplary graph state, the result of this simulation is shown in Fig.~\ref{fig_results_all}. The considered graph has only two non-isomorphic colorings. When choosing the independent sets $A$ according to the first coloring, the best purification protocol is shown in Fig.~\ref{fig_results_all}(a). For the other coloring, the result is shown in Fig.~\ref{fig_results_all}(b). The comparison of the two simulation results reveals interesting differences. In Fig.~\ref{fig_results_all}(a), there are clear structures that indicate for which noise parameters which order of the $k$ purifications is best. In Fig.~\ref{fig_results_all}(b), the result appears noisy over the entire parameter regime because there are always several purification protocols with identical performance. This degeneracy occurs because the different colors are indistinguishable in the graph coloring from Fig.~\ref{fig_results_all}(b). When swapping for instance the blue with the red color, one obtains exactly the same graph and exactly the same coloring. Therefore, there are only a few distinct cascaded purification protocols which are distinguished by the gray scale in Fig.~\ref{fig_results_all}(b). In the low noise regime, for example, it is best to choose the set $A$ according to a different color for each of the three purification rounds. Due to the indistinguishability of the colors, there are $3!=6$ orderings of the colors that correspond to the identical purification protocol.

Fig.~\ref{fig_results_all}(c) shows the best fidelity that can be reached after all three purification rounds, where we take the optimum over the data from both colorings. Fig.~\ref{fig_results_all}(d) is the corresponding optimum for the fraction of removed infidelity. It turns out that the choice of the coloring only has a small effect of plus or minus a few percent. To illustrate in which regime each choice of coloring is best, we show in Fig.~\ref{fig_results_all}(e), the difference between the best fidelities from (a) minus the best fidelities from (b). Fig.~\ref{fig_results_all}(f) shows the analogous plot for the fraction of removed infidelity.

\section{Multiplexing overhead}
\label{sec_multiplex}
When type-I fusions are employed, purification protocols only succeed with finite probability even when the input states have fidelities close to one. To compensate for the limited success probability, multiplexing will be required, and we discuss a practical case of spatial multiplexing. We will assume high-fidelity input states, neglecting the contribution of finite fidelities to the success probabilities.

In fusion-based quantum computing, many identical high-fidelity resource states are required simultaneously~\cite{Bartolucci2021}. Say one requires at least $n_0$ purified resource states in a certain time interval, and the probability $p_t$ that we receive enough states should be very high. Obtaining fewer resource states will not necessarily cause the fusion-based computation to fail. However, missing resource states correspond to (correlated) photon loss with loss thresholds for fault-tolerance in the few percent regime~\cite{Bartolucci2021}. Therefore, we will have to choose the number $n$ of redundant purifications such that $\sum_{k=0}^{n_0-1}p_s^k(1-p_s)^{n-k}\binom{n}{k}\leq 1-p_t$, where $p_s$ is the success probability of the purification. In the limit of very many required resource states (large $n_0$), the required number $n$ will only be slightly higher than $n_0/p_s$.

\begin{figure}
\includegraphics[width=1.0\linewidth]{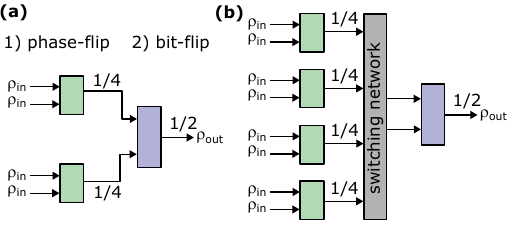}
\caption{\label{fig_multiplex}\textbf{(a)} Two cascaded purification rounds for a $3$-photon GHZ state. The first purification layer filters phase-flip errors and succeeds with a probability of up to $p_s^{(p)}=1/4$, depending on the input fidelity. The second purification layer filters bit-flip errors and succeeds with a probability of up to $p_s^{(b)}=1/2$. \textbf{(b)} The two-layer purification is performed with more input states such that multiplexing is possible after the first layer, where a switching network collects two successfully purified states from the first layer. This in-between multiplexing can reduce the overall overhead.}
\end{figure}

Often more than one purification round is required to achieve the target fidelities. Assume $d$ rounds of cascaded purification where $c_i$ represents the average number of imperfect input states to obtain one purified state at layer $i$ ($d+1$ layers for $d$ rounds). We then have $c_{i+1}=2c_i/p_i, c_1=1$, where $p_i$ is the probability that purification succeeds (Ref.~\cite{Lee2023} uses a similar recursion in the context of graph state generation). For $d$ rounds of cascaded purification, the average number of input states to obtain each output state is then given by $2^d/\prod_{i=1}^{d}p_i$. This formula implies a divide-and-conquer approach which assumes that, if a purification on the $i$th layer fails, it can be repeated, without affecting other states at the same layer. This means that the other states at the same layer may need to be stored in an optical fiber until the failed purification is repeated successfully. Such multiplexing in time is not ideal as it increases photon loss. Instead, one could use a static cascaded purification device like the one shown in Fig.~\ref{fig_multiplex}(a) and multiplex many devices in space. However, this device only succeeds in producing a purified output if all purifications of the binary tree succeed. The average number of input states to obtain each output state would be $2^d/\prod_{i=1}^{d}p_i^{2^{d-i}}$ and thus scale super-exponentially with $d$. This issue can be avoided by performing multiplexing in space at every layer as illustrated in Fig.~\ref{fig_multiplex}(b). By generating more than the minimum number of states at the initial layer, one ensures that enough states will be present at the next layer with high probability. These states are then routed by a switching network to the next layer of purification setups. From layer $i$ to layer $i+1$, the number of states will be on average reduced by a factor of $p_i/2$. Asymptotically, the number of input states to obtain each output state is therefore again $2^d/\prod_{i=1}^{d}p_i$, like in the case where multiplexing in time is considered.

As an example, we discuss the setups in Fig.~\ref{fig_multiplex}. There are two cascaded purification rounds, with a first purification for phase-flip and a second purification for bit-flip errors. The scheme in Fig.~\ref{fig_multiplex}(a) uses the minimum number of four input states. As soon as a single purification round does not succeed, the device does not output a state, and thus the overall success probability is the product of the individual success probabilities: $p_s^{(b)}\cdot p_s^{(p)}\cdot p_s^{(p)} = 1/32$, where $p_s^{(b)}=1/2, p_s^{(p)}=1/4$ are the probabilities of successful bit- and phase-flip purification for a three-qubit GHZ state. The second device in Fig.~\ref{fig_multiplex}(b) does twice as many purifications in the first layer and thus needs twice as many input states. Successfully purified states are routed via a switching network towards the second purification layer. By using more ($n=4$) purifications in the first layer, the probability of at least two phase-flip purified states being available at the second layer is $p_2=1-(1-p_s^{(p)})^n-n\cdot p_s^{(p)}(1-p_s^{(p)})^{n-1}$, which leads to an overall success probability of $p_s=p_2\cdot p_s^{(b)}\simeq 0.13$. This probability is much higher than the success probability $1-(1-1/32)^2\simeq 0.06$ that would be achieved by multiplexing two devices from Fig.~\ref{fig_multiplex}(a) and thus using the same number of input states. This example illustrates that multiplexing at intermediate layers can help reduce overhead.

We aim to minimize the overhead in terms of number of photons consumed to build a certain resource state, and we compare the overhead to an all-linear-optics generation of a 3-qubit GHZ-state. We assume that an all-linear-optics scheme could generate the state with a fidelity that is as good as what we obtain from two purification (bit-flip and phase-flip) using 3-qubit GHZ-states generated by quantum emitters~\cite{Meng2023b, Lindner2009}. This is a generous assumption from the perspective of quantum emitters: experimentally, the fidelities of three-qubit GHZ-states, generated probabilistically with the standard linear optics scheme, reached $F=0.82$~\cite{Maring2024, Cao2025}. Compared to deterministic entanglement generation, this value is similar to the fidelities achieved with solid-state quantum emitters~\cite{Meng2023b, Huet2024} and is clearly below the fidelities achieved with atoms~\cite{Thomas2022}. The linear optics scheme succeeds in generating the 3-qubit GHZ-state with probability $1/32$ using $6$ input photons~\cite{Li2015, Maring2024} and thus requires on average $6\cdot 32=192$ photons to produce the 3-qubit GHZ-state. For purifying emitter-generated 3-qubit GHZ-states, we use the scheme in Fig.~\ref{fig_multiplex}(b) with $n$ initial phase-flip purification rounds. The overhead, i.e. the average number of photons required to make one purified state, is then $6n/p_s$, where the factor $6$ comes from the fact that we need two input states per purification with three qubits each, and $p_s$ is the overall success probability. For the considered two-stage purification round in Fig.~\ref{fig_multiplex}(b), we find that the overhead is minimized at $n=7$, leading to an overhead of $151$ photons used on average. Instead, performing bit-flip purifications before one phase-flip purification, we find an optimum at $n=4$ with an overhead of $141$ photons. Instead, one could use multiplexing in time, where one would generate phase-purified states one-by-one until exactly two such states are present for the subsequent bit-flip purification. This would require storing the first phase-flip purified state in a delay line until the second one is generated. The overhead to obtain two phase-flip purified states would be $2\times6/p_s^{(p)}$. Taking these two states, the subsequent bit-flip purification round succeeds with $p_s^{(b)}$ and the overall overhead would thus be $2\times6/p_s^{(p)}/p_s^{(b)}=96$. In all cases, the overhead is below the $192$ photons required in an all-linear-optics 3-qubit GHZ-state generation~\cite{Maring2024}.

Although this estimation does not include the effect of finite fidelities on the probability of successful purification, it shows that the overhead can be lower than in an all-linear-optics scheme even if two purification rounds are executed. If just one purification round is sufficient, the overhead is much lower. Also note that the above estimations do not consider the efficiency of the used photon sources. If the all-linear-optics schemes used photons from single photon sources, we may assume that the efficiencies are similar to the case where entangled photonic states are generated deterministically. If photons are generated via down conversion the efficiency may be different, but then there would be an additional overhead from the low probability of generating single photons.

\section{Summary and Outlook}
We have developed bit- and phase-flip purification schemes for photonic graph states using type-I fusions~\cite{Browne2005} instead of deterministic entangling gates. Our schemes apply to GHZ states, more general CSS states, and arbitrary graph states. For certain CSS states, we have also shown how to simultaneously purify bit- and phase-flip errors but not Pauli $Y$ errors. Numerically, we determined the purification protocols that give the highest fidelities for various graph states. Furthermore, we outlined a way to reduce multiplexing overhead in cascaded purification.

We suggest two directions for further analysis: first, we assume that the purification itself is close to error-free. This assumption is currently justified, since recent infidelities of photonic graph states generated by quantum emitters~\cite{Meng2023b, Huet2024} are much higher than error rates in state-of-the-art linear optics~\cite{Alexander2025}. Furthermore, error rates of resource states are likely enhanced when a highly adaptive graph state generation is used~\cite{Staudacher2026, Duan2005}. For long term applications of graph states in fusion based quantum computation~\cite{Bartolucci2021} it is, however, essential that the fidelity of the final states is very high so that the assumption of error-free purification may no longer be applicable. This is particularly relevant after a few purification rounds, where the initial infidelity has been suppressed so that the final fidelity is limited by the errors of the circuit. Imperfect purification circuits~\cite{Goyal2006} should therefore be investigated in detail. Second, we have only considered errors, but the combined effect of errors and photon loss should be investigated. We take a first look at this in Appendix~\ref{sec_loss}, where we show that to lowest order the effect of loss is not increased by the considered bit-flip purification, whereas it is doubled for the phase flip purification protocol.

While some further analysis may be required, our results indicate that purification could become a key ingredient for applications such as fusion-based quantum computing~\cite{Bartolucci2021} where high-fidelity photonic entangled states are required. As an additional layer of post-selection, our resource state purification could also be added to architectures where parts of the fusion network are post-selected based on the outcome of its type-II fusions~\cite{Birchall2026}. Some of the graph states analyzed here have already been built experimentally~\cite{Huet2026, Meng2023b} and remarkable progress has been made in optimizing linear optics~\cite{Alexander2025}. These achievements make a near-term realization of our purification protocols conceivable.

\section{Acknowledgement}
We thank Oliver Sandberg and Love Pettersson for useful discussions. The authors acknowledge financial support from Danmarks Grundforskningsfond (DNRF 139, Hy-Q Center for Hybrid Quantum Networks), the Novo Nordisk Foundation (Challenge project ”Solid-Q”), and Danmarks Innovationsfond (IFD1003402609, FTQP). A.A.C. acknowledges funding from UK EPSRC (EP/S023607/1). M.L.C acknowledges funding from Danmarks Innovationsfond (Grant No.~4298-00011B). S.P. acknowledges funding from VILLUM FONDEN (MapQP, No. VIL60743), the European Research Council (ERC StG ASPEQT, No. 101221875), and funding support from the NNF Quantum Computing Programme. The authors thank Amazon Web Services (AWS) for the free computing time used for the numerical simulations performed for this project.

\appendix

\section{Path encoding}
\label{sec_dual_rail}
\begin{figure}
\includegraphics[width=1.0\linewidth]{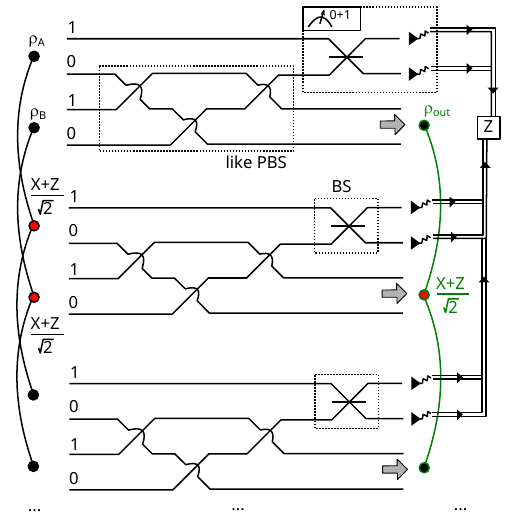}
\caption{\label{fig_setup_dual_rail} Purification setup for path encoded qubits, where a photon in one of two modes $0,1$ represents the computational state. The polarizing beam splitters in Fig.~\ref{fig_setups}(a) are replaced by mode swaps of the computational $\ket{0}$ state. Measurements in the rotated $\ket{0}+\ket{1}$ basis are implemented by placing a balanced beam-splitter (BS) before the detectors.}
\end{figure}

In Figs.~\ref{fig_setups} and \ref{fig_time_bin}, we show setups for the purification of polarization and time-bin-encoded states, respectively. Path encoding is another approach that is widely used in photonic quantum computing~\cite{Alexander2025}. In Fig.~\ref{fig_setup_dual_rail} we show a corresponding purification setup for this encoding scheme.

\section{CSS purification success probability}
\label{CSS_probabiliy}
In the main text, we have shown that the success probability for bit-flip purification of a CSS state is bounded by $1/2^{n_x}$, where $n_x$ is the number of independent $X$-type stabilizers, i.e. the rank of the stabilizer matrix $H_x$. In this section, we give an alternative explanation considering a CSS state in the Schr{\"o}dinger picture.

Let $z_i$ represent the $i$th row of the $Z$-type stabilizer generator matrix $H_z$. We assume without loss of generality (see below) that all stabilizers have positive sign, i.e. $a_x=0$ and $a_z=0$. We write the CSS state as $\ket{\Psi}=\sum_{x\in \mathbb{F}_2^n}\alpha_x\ket{x}$, where $\ket{x}$ is the binary representation of $x$. By definition, we have $\forall_{z_i \in H_z} Z^{\otimes z_i}\ket{\Psi}=\ket{\Psi}$ and therefore (inserting the decomposition of $\ket{\Psi}$) $\forall_{z_i \in H_z} \forall_{x \in\mathbb{F}_2^n}\, \alpha_x = (-1)^{x\cdot z_i}\alpha_x$. For all non-zero coefficients $\alpha_{x}$, this implies $x\cdot z_i=0$ meaning that $x$ is in the orthogonal/dual space~\cite{Macwilliams1977} of the $H_z$ rowspace. Since $X$- and $Z$-type stabilizers must commute and we have a full set of $n_x+n_z=n$ stabilizers, this orthogonal space is exactly the rowspace of $H_x$~\cite{Chen2004}~\footnote{For a CSS state, the $X$-type stabilizers thus define the $Z$-type stabilizers and vice versa.}. Therefore, $\alpha_x$ is only non-zero if the binary pattern of $x$ represents an $X$-type stabilizer. Furthermore, taking into account the $X$-type stabilizers, we have $\forall_{x_i \in H_x} X^{\otimes x_i}\ket{\Psi}=\ket{\Psi}$ and thus $\forall_{x_i \in H_x} \forall_{x \in\mathbb{F}_2^n}\, \alpha_x = \alpha_{x\oplus x_i}$, where $\oplus $ represents a bit-wise XOR operation. Therefore, the CSS state has the form~\cite{Nielsen2010}
\begin{equation}
    \label{eq_css_schr}
    \frac{1}{\sqrt{2^{n_x}}}\sum_{x_i \in H_x}\ket{x_i},
\end{equation}
which tells us where the $1/2^{n_x}$ success probability in the purification comes from: there are $2^{n_x}$ mutually orthogonal terms in Eq.\eqref{eq_css_schr} and if we send two such states into a transversal type-I fusion, the probability that the qubit polarizations coincide on all type-I fusions (criterion for success) is thus $1/2^{n_x}$. For different CSS states where not all stabilizers have positive sign, i.e. $a_x\neq 0$ or $a_z\neq0$, this conclusion does not change. Any sign-flip in the stabilizer tableau corresponds to Pauli matrices being applied to the state. In Eq.\eqref{eq_css_schr}, any Pauli $Z$-operator only adds $-1$ prefactors while any Pauli $X$-operator adds a constant offset to all binary pattern but neither operation changes the number of $2^{n_x}$ mutually orthogonal states.

\section{Effect of photon loss}
\label{sec_loss}
In the main text, we have considered purification protocols in the presence of Pauli errors. Another issue for photonic quantum computing is photon loss. Despite recent progress in improving the efficiency of quantum-emitter-based photon sources~\cite{Chu2017,Tomm2021,Ding2025,Loredo2025} and the development of such sources in telecom bands~\cite{Albrechtsen2025,Joos2024}, losses still remain high. Therefore, we consider here the effect of photon loss in the GHZ purification protocols from Section~\ref{sec_ghz}. We find that, to first order, loss is not increased by bit flip purification, whereas it roughly doubles for phase flip purification.

\subsection{Bit-flip purification with loss}
We assume that each photon from the two input states is lost with probability $p_l$. The density matrix of an initial unpurified state is then given by
\begin{equation}
    \rho=(1-p_l)^n\ket{\Phi_0^+}\bra{\Phi_0^+} + \sum_{l=1}^{2^n-1}(1-p_l)^{n-q_l}p_l^{q_l}\ket{\Phi_0^+}\bra{\Phi_0^+}_{l=\emptyset},
\end{equation}
where $q_l$ is again the sum of the bitwise representation of $l$ as before and $l=\emptyset$ indicates that all photons are lost for which the bitwise representation of $l$ is non-zero.

For the cross-terms between two input states, we consider three cases separately: (1) both input states are not lossy. (2) Only one of the input states is lossy, but not all of its qubits are lost. (3) Both input states have lossy qubits or all input qubits of one part are lost. In case (1), we obtain the desired state $\ket{\Phi_0^+}$ with probability $1/2$ and the overall probability is thus $\frac{1}{2}(1-p_l)^{2n}$. Case (2) corresponds to an input state of the form $\ket{\Phi_0^+}_{A, l=\emptyset}\otimes\ket{\Phi_0^+}_B = \frac{1}{\sqrt{2}}\left(\ket{H..H}_{l=\emptyset}+\ket{VV..V}_{l=\emptyset}\right)_A \otimes \frac{1}{\sqrt{2}}\left(\ket{HH..H}+\ket{VV..V}\right)_B$ (case of loss in the system $A$). By the effect of the PBSs this becomes:
\begin{align}
    &\frac{1}{2}\ket{HH..H}_A\ket{HH..H}_{B,l=\emptyset}\label{eq_after_pbs__loss_1}\\
    +&\frac{1}{2}\ket{VV..V}_{A,l=\emptyset}\ket{VV..V}_B\label{eq_after_pbs__loss_2}\\
    +&\frac{1}{2}\ket{HH..H}_A\ket{VV..V}_{A,l=\emptyset}\label{eq_after_pbs__loss_3}\\
    +&\frac{1}{2}\ket{HH..H}_{B, l=\emptyset}\ket{VV..V}_B\label{eq_after_pbs__loss_4}.
\end{align}
Note that only the first term leads to a detection pattern that is not filtered out, leading to the state $\ket{H}^{\otimes n}_{l=\emptyset}$ with probability $1/4$. If the lossy qubits are in the other input state $B$, we would instead obtain $\ket{V}^{\otimes n}_{l=\emptyset}$. The overall probability of obtaining one particular such state is therefore $\frac{1}{4}(1-p_l)^{2n-q_l}p_l^{q_l}$. In case (3), there is only one case in which the state is not filtered out by the detection pattern and that is the case in which the loss pattern $l$ in system $A$ is the complement of the loss pattern $l'$ in system $B$, i.e. each qubit is either lost in system $A$ or $B$ but not in both. This case has a probability of $(1-p_l)^np_l^n$. With probability $1/4$ (only the corresponding term in Eq.~\eqref{eq_after_pbs__loss_3} is not filtered by post-selection), one click per detector is measured and the output state is completely empty, i.e.  $\ket{\emptyset}^{\otimes n}$. Only in the two cases where every qubit in $A$ or $B$ is lost, there is an additional contribution from the terms in Eq.~\eqref{eq_after_pbs__loss_1} or Eq.~\eqref{eq_after_pbs__loss_2} and the terms are thus not filtered with probability $1/2$.

Overall, we obtain the density matrix (up to some normalization factor corresponding to the probability of one click on each detector)
\begin{align}
    &\frac{1}{2}(1-p_l)^{2n}\ket{\Phi_0^+}\bra{\Phi_0^+}\notag\\
    +&\frac{1}{4}\sum_{l=1}^{2^n-2}(1-p_l)^{2n-q_l}p_l^{q_l}\left[\ket{H}^{\otimes n}\bra{H}^{\otimes n} + \ket{V}^{\otimes n}\bra{V}^{\otimes n}\right]_{l=\emptyset}\notag\\
    +&(1-p_l)^np_l^n\left(\frac{1}{4}(2^n-2)+\frac{1}{2}2\right)\ket{\emptyset}^{\otimes n}\bra{\emptyset}^{\otimes n},
\end{align}
where the three lines correspond to the three cases considered above. The relative fraction of the desired term in the first line is given by
\begin{align}
    \label{eq_loss_bit_GHZ_fraction}
   & (1-p_l)^{2n} / [(1-p_l)^{2n}\notag \\
    &+ (1-p_l)^n(1-(1-p_l)^n-p_l^n) \notag\\
    &+ p_l^n(1-p_l)^n\left(\frac{2^n-2}{4}+1\right)].
\end{align}
In first-order Taylor expansion around $p_l=0$ this corresponds to $1-np_l$ which coincides with the corresponding first-order Taylor expansion of $(1-p_l)^n$, which is the fraction of the non-lossy state in the input. Therefore, there is no amplification of loss if $p_l$ is sufficiently small. Furthermore, Eq.~\eqref{eq_loss_bit_GHZ_fraction} coincides with $(1-p_l)^n$ in the asymptotic regime of large $n$. We also find numerically that the deviation between Eq.~\eqref{eq_loss_bit_GHZ_fraction} and $(1-p_l)^n$ is very small in the range relavant for fusion-based quantum computing ($0\leq p_l \leq 0.1$)~\cite{Bartolucci2021, Chan2024}. All of this indicates that the bit-flip purification of GHZ states does not have a problematic effect on the loss of the output state.

\subsection{Phase-flip purification with loss}
We now consider loss in the phase-flip purification protocol. In the presence of loss, we have an initial density matrix given by
\begin{equation}
\label{eq_phase_loss}
    \rho = \sum_{l=0}^{2^n-1}(1-p_l)^{n-q_l}p_l^{q_l} H^{\otimes n}\ket{\Phi_0^+}\bra{\Phi_0^+}_{l=\emptyset}H^{\otimes n},
\end{equation}
with $H^{\otimes n}\ket{\Phi_0^+} = \frac{1}{\sqrt{2^{n-1}}}\sum_{k=0}^{2^n-1}E(q_k)\ket{k}$ following from Eq.~\eqref{eq_even_general}. We again distinguish three cases. In the first case, there is no loss in either of the two input states (probability $(1-p_l)^{2n}$). In this case, we obtain $H^{\otimes n}\ket{\Phi_0^+}$ with probability $\frac{1}{2^{n-1}}$ (see Section~\ref{sec_phase_ghz}) and the overall probability of obtaining $H^{\otimes n}\ket{\Phi_0^+}$ is thus $\frac{1}{2^{n-1}}(1-p_l)^{2n}$. In the second case, only one of the two input states is lossy (can be completely lost). Say we have losses in system $A$ with a loss pattern $l$ and no loss in $B$, corresponding to a probability $(1-p_l)^{2n-q_l}p_l^{q_l}$. For every term $\ket{k}_{l=\emptyset}$ in Eq.~\eqref{eq_phase_loss} (for system $A$) there is exactly one term $\ket{\Tilde{k}}$ in Eq.~\eqref{eq_even_general} (for system $B$) for which the state is not filtered by post-selection. This is the term $\ket{\Tilde{k}}$ for which the bitwise representation of $\Tilde{k}$ coincides with $k$ on all non-lost qubits and is $0$ ($H$-polarization) on all qubits that are lost in $A$. The probability that the state is not filtered is therefore again $\frac{1}{2^{n-1}}$. The state obtained in this case is $H^{\otimes n}\ket{\Phi_0^+}_{l=\emptyset}$. In the third case, there is at least one loss in both input states. Two loss patterns $l, l'$ appear with probability $\alpha_{l, l'}=(1-p_l)^{n-q_l}p_l^{q_l}(1-p_{l'})^{n-q_{l'}}p_l^{q_{l'}}$. This case is always filtered by post-selection if there is a loss on the same qubit of both input states (i.e. if $l\&l' \neq 0$). If $l\&l' = 0$, there is no filtering if (1) two terms $\ket{k}_{l=\emptyset}$ and $\ket{k'}_{l'=\emptyset}$ coincide on all non-lost qubits, and (2) one qubit in $k$ ($k'$) being lost is compensated by the corresponding qubit in $k'$ ($k$) having the value that leads to the detection of one photon. The probability  for not being filtered by post-selection is therefore again $\frac{1}{2^{n-1}}$. Note that in the third case, more than one photon is lost and therefore the overall probability for this case is $\mathcal{O}(p_l^2)$, which does not contribute to the leading order.

Putting the above considerations together, we obtain a density matrix $\rho$ which, up to normalization, is given by
\begin{align}
    &H^{\otimes n}\rho H^{\otimes n} =\notag \\
    &\frac{1}{2^{n-1}}(1-p_l)^{2n}\ket{\Phi_0^+}\bra{\Phi_0^+}\notag\\
    +&\frac{2}{2^{n-1}}\sum_{l=1}^{2^n-1}(1-p_l)^{2n-q_l}p_l^{q_l}\ket{\Phi_0^+}\bra{\Phi_0^+}_{l=\emptyset}\notag\\
    +&\frac{1}{2^{n-1}}\sum_{l=1}^{2^n-1}\sum_{l'=1, l'\& l = 0}^{2^n-1}\alpha_{l, l'}\ket{\Phi_0^+}\bra{\Phi_0^+}_{l\oplus l'=\emptyset}.
\end{align}
The relative fraction of the desired term in the first line is given by $(1-p_l)^{2n} / [(1-p_l)^{2n} + 2(1-p_l)^n(1-(1-p_l)^n) + \sum_{l=1}^{2^n-1}\sum_{l'=1, l'\& l = 0}^{2^n-1}\alpha_{l, l'}]$. In first-order Taylor expansion around $p_l=0$ this corresponds to $1-2np_l$ which doubles the loss for small $p_l$. Furthermore, we find numerically that the relative fraction of the desired term is very close to the probability $(1-p_l)^{2n}$ of no loss in both input states.

\bibliography{lit}

\end{document}